\documentclass[11pt]{article}\def\today{March 2, 2006}
\usepackage{amssymb}
\usepackage{amsmath}
\usepackage{theorem}
\usepackage{latexsym}
\usepackage{graphicx}
\usepackage[all]{xy}
\reversemarginpar
\theoremheaderfont{\bf} %\theoremstyle{break}
\theorembodyfont{\sl}
\newtheorem{theorem}{Theorem}[section]
{\theorembodyfont{\rm} \newtheorem{definition}[theorem]{Definition}}
{\theorembodyfont{\rm} \newtheorem{exa}[theorem]{Example}}
{\theorembodyfont{\rm} \newtheorem{remark}[theorem]{Remark}}
\newtheorem{proposition}[theorem]{Proposition}
\newtheorem{corollary}[theorem]{Corollary}
\newtheorem{conj}[theorem]{Conjecture}
{\theorembodyfont{\rm} \newtheorem{ass}[theorem]{General Assumption}}
\newtheorem{lemma}[theorem]{Lemma}
{\theorembodyfont{\rm}}
{\theorembodyfont{\rm}}
\newenvironment{proof}{{\sc Proof:}}{\mbox{}\hfill$\Box$\par}
\newcommand{\eqnref}[1]{~\mbox{$(${\rm \ref{#1}}$)$}}

\newcommand{\junk}[1]{}
\newcommand{\DS}{\displaystyle}
\newcommand{\TS}{\textstyle}
\newcommand{\N}{{\mathbb N}}

\newcommand{\F}{{\mathbb F}}
\newcommand{\C}{{\mathbb C}}

\newcommand{\cC}{{\mathcal C}}

\newcommand{\cF}{{\mathcal F}}
\newcommand{\cG}{{\mathcal G}}

\newcommand{\cM}{{\mathcal M}}
\newcommand{\cI}{{\mathcal I}}

\newcommand{\cJ}{{\mathcal J}}
\newcommand{\cP}{{\mathcal P}}
\newcommand{\cH}{{\mathcal H}}

\newcommand{\HH}{\mbox{\rm\textbf H}}
\newcommand{\cCconst}{\mbox{${\mathcal C}_{\,\text{\rm const}}$}}
\newcommand{\cCperpconst}{\mbox{$\widehat{\mathcal C}_{\,\text{\rm const}}$}}
\newcommand{\CC}{\mbox{$C_{_{\scriptstyle\mathcal C}}$}}
\newcommand{\CCperp}{\mbox{$C_{_{\scriptstyle\widehat{\mathcal C}}}$}}
\newcommand{\rank}{\mbox{\rm rank}\,}

\newcommand{\im}{\mbox{\rm im}\,}

\newcommand{\wt}{\mbox{\rm wt}}
\newcommand{\we}{\mbox{\rm we}}
\newcommand{\cwe}{\mbox{\rm cwe}}

\newcommand{\spann}[1]{\mbox{$\langle{#1}\rangle$}}

\newcommand{\Smalltwomat}[2]{\mbox{\large{\mbox{$\genfrac{(}{)}{0pt}{1}{#1}{#2}$}}}}
\newcommand{\Smallfourmat}[4]{\mbox{\scriptsize{$\begin{pmatrix}{#1}&{\!\!\!#2}\\{#3}&{\!\!\!#4}\end{pmatrix}$}}}
\newcounter{abc}
\newcounter{def}

\newenvironment{alphalist}{\begin{list}{(\alph{abc})\hfill}{\usecounter{abc}
     \topsep-1.4ex \labelwidth.6cm \leftmargin.6cm \labelsep0cm
     \rightmargin0cm \parsep0ex \itemsep.6ex
     \partopsep1.6ex}}{\end{list}}
\newenvironment{arabiclist}{\begin{list}{(\arabic{abc})\hfill}{\usecounter{abc}
     \topsep-1.4ex \labelwidth.7cm \leftmargin.7cm \labelsep0cm
     \rightmargin0cm \parsep0ex \itemsep.6ex
     \partopsep1.6ex}}{\end{list}}

\title{On the MacWilliams Identity for Convolutional Codes}
\date\today
\author{Heide Gluesing-Luerssen$^*$, Gert Schneider\footnote{
       University of Groningen, Department of Mathematics, P.~O.~Box 800,
       9700 AV Groningen, The Netherlands; gluesing@math.rug.nl, schneider@math.rug.nl}
       }
      
\begin{document}
\maketitle
{\bf Abstract:} 
The adjacency matrix associated with a convolutional code collects in a detailed manner information about the weight distribution of the code.
A MacWilliams Identity Conjecture, stating that the adjacency matrix of a code fully determines the adjacency matrix of the dual code, will be formulated, and an explicit formula for the transformation will be stated.
The formula involves the MacWilliams matrix known from complete weight enumerators of block codes. 
The conjecture will be proven for the class of convolutional codes where either the code itself or its dual does not have Forney indices bigger than one.
For the general case the conjecture is backed up by many examples, and a weaker version will be established. 



%%%%%%%%%%%%%%%%%%%%%%%%%%%%%%%%%%%%%%%%%
\section{Introduction}\label{S-Intro}
\setcounter{equation}{0}
%%%%%%%%%%%%%%%%%%%%%%%%%%%%%%%%%%%%%%%
Two of the most famous results in block code theory are MacWilliams' Identity Theorem and Equivalence Theorem~\cite{MacW62},~\cite{MacW63}. 
The first one relates the weight enumerator of a block code to that of its dual code. The second one states that two isometric codes are monomially equivalent. 
The impact of these theorems for practical as well as theoretical purposes is well-known, see for instance \cite[Ch.~11.3, Ch.~6.5, Ch.~19.2]{MS77} or the classification of constant weight codes in \cite[Thm.~7.9.5]{HP03}.

The paramount importance of the weight function in coding theory makes an understanding of weight enumerators, isometries, and, more explicitly, possible versions of the MacWilliams Theorems a must for the analysis of any class of codes.
For instance, after realizing the relevance of block codes over finite rings, both theorems have seen generalizations to this class of codes, see for instance~\cite{Wo99} and~\cite{DiLP04}.
For convolutional codes the question of a MacWilliams Identity Theorem has been posed already about~$30$ years ago. 
In~$1977$ Shearer/McEliece~\cite{SM77} considered the weight enumerator for convolutional codes as introduced by Viterbi~\cite{Vi71}. 
It is a formal power series in two variables counting the number of irreducible (``atomic'') codewords of given weight and length; for the coding-theoretic relevance see, e.~g., 
\cite[Sec.~VII]{Vi71} and \cite[Sec.~4.3]{JoZi99}.
Unfortunately, a simple example in~\cite{SM77} made clear that a MacWilliams Identity 
does not exist for these objects.
A main step forward has been made in $1992$ when Abdel-Ghaffar~\cite{Ab92} considered a more refined weight counting object: the weight enumerator state diagram.
For unit constraint-length codes he derives a MacWilliams Identity in form of a list of separate formulas relating the labels of this diagram to those of the dual code.

In this paper we will present a MacWilliams Identity for the class of convolutional codes where either the code or its dual does not have Forney indices bigger than one. 
Duality of codes will be defined in the standard way based on the vanishing of the canonical bilinear form on~$\F[z]^n$.
Our result generalizes not only the block code case, but also Abdel-Ghaffar's transformation for unit constraint-length codes.
We will show in Section~\ref{S-AbGh} that the list of identities given in~\cite{Ab92} can be written in closed form just like in our MacWilliams Identity.
In addition to the result just mentioned we will also formulate an explicit conjecture on a MacWilliams Identity for all classes of convolutional codes.
It is backed up by a wealth of examples, and a weaker version will be proven. 

The weight counting object in our considerations is the so-called adjacency matrix of the encoder.
This matrix has been discussed in detail by McEliece~\cite{McE98a}, but appears already in different notations earlier in the literature. 
Indeed, one can show that it basically coincides with the labels of the weight enumerator state diagram as considered in~\cite{Ab92}.
The adjacency matrix is defined via a state space description of the encoder as introduced in~\cite{MaSa67}.
In this sense our approach follows a series of papers where system theoretic methods have been used successfully in order to investigate convolutional codes, see for instance~\cite{Ro01},~\cite{RoYo99},~\cite{FaZa01}, and~\cite{HRS05}.
The matrix is labeled by the set of all state pairs, and each entry contains the weight enumerator of all outputs associated with the corresponding state pair. 
The whole matrix contains considerably more detailed information about the code than the weight enumerator discussed above. 
Indeed, it is well-known \cite{McE98a},~\cite{GL05p} how to derive the latter from the adjacency matrix.
Furthermore, in~\cite{GL05p} it has been shown that, after factoring out the group of state space isomorphisms, the adjacency matrix turns into an invariant of the code, called the generalized adjacency matrix. 

The main outline of our arguments is as follows.
In the next section we will introduce two block codes canonically associated with a convolutional code.
They are crosswise dual to the corresponding block codes of the dual convolutional code.
Later on this fact will allow us to apply the MacWilliams Identity for block codes suitably.
Indeed, in Section~\ref{S-adjmatrix} we will introduce the adjacency matrix~$\Lambda$ and show that its nontrivial entries are given by the weight enumerators of certain cosets of these block codes.
The main ingredient for relating~$\Lambda$ with the adjacency matrix of the dual 
will be a certain transformation matrix~$\cH$ as it also appears for the {\em complete\/} weight enumerator of block codes.
This matrix will be studied in Section~\ref{S-MacWMat}, and a first application to the adjacency matrix will be carried out.
In Section~\ref{S-Duality} we will be able to show our main results.
Firstly, we prove that entrywise application of the block code MacWilliams Identity for the matrix $\cH\Lambda^t\cH^{-1}$ will result in a matrix that up to reordering of the entries coincides with the adjacency matrix of the dual code. 
Secondly, for codes where the dual does not have Forney indices bigger than one we will show that the reordering of the entries comes from a state space isomorphism.
As a consequence, the resulting matrix is indeed a representative of the generalized 
adjacency matrix of the dual code. 
This is exactly the contents of our MacWilliams Identity Theorem.

We end the introduction with recalling the basic notions of convolutional codes.
Throughout this paper let 
\begin{equation}\label{e-Fq}
   \F=\F_q\text{ be a finite field with~$q=p^s$ elements where $p$ is prime and $s\in\N$}. 
\end{equation}
A {\em $k$-dimensional convolutional code of length\/} $n$ is a submodule~$\cC$ of $\F[z]^n$ of the form 
\[
    \cC=\im G:=\{uG\,\big|\, u\in\F[z]^k\}
\]
where~$G$ is a {\em basic\/} matrix in $\F[z]^{k\times n}$, i.~e. there exists some matrix $\tilde{G}\in\F[z]^{n\times k}$ such that $G\tilde{G}=I_k$. 
In other words,~$G$ is noncatastrophic and delay-free.
We call $G$ an {\em encoder\/} and the number 
$\delta:=\max\{\deg\gamma\mid\gamma\text{ is a $k$-minor of }G\}$ is said to be the 
{\sl degree\/} of the code~$\cC$.
A code having these parameters is called an $(n,k,\delta)$ code.
A basic matrix $G\in\F[z]^{k\times n}$ with rows $g_1,\ldots, g_k\in\F[z]^n$ is said to be 
{\em minimal\/} if $\sum_{i=1}^k\deg (g_i)=\delta$.
For characterizations of minimality see, e.~g., \cite[Main~Thm.]{Fo75} or \cite[Thm.~A.2]{McE98}.
It is well-known~\cite[p.~495]{Fo75} that each convolutional code~$\cC$ admits a minimal encoder~$G$.
The row degrees $\deg g_i$ of a minimal encoder~$G$ are uniquely determined up to ordering and are called the {\em Forney indices\/} of the code or of the encoder. 
It follows that a convolutional code has a constant encoder matrix if and only if the degree is zero. 
In that case the code can be regarded as a block code.

The weight of convolutional codewords is defined straightforwardly.
We simply extend the ordinary {\em Hamming weight\/}
$\wt(w_1,\ldots,w_n):=\#\{i\mid w_i\not=0\}$ defined on~$\F^n$
to polynomial vectors in the following way.
For $v=\sum_{j=0}^Nv^{(j)}z^j\in\F[z]^n$, where $v^{(j)}\in\F^n$, we put the {\em weight\/} of~$v$ to be $\wt(v)=\sum_{j=0}^N\wt(v^{(j)})$.

Finally we fix the following notions.
For $\delta>0$ we will denote by $e_1,\ldots,e_\delta$ the unit vectors in $\F^\delta$. 
For any matrix $M\in\F^{a\times b}$ we denote by $\im M:=\{uM\mid u\in\F^a\}$ 
and $\ker M:=\{u\in\F^a\mid uM=0\}$
the image and kernel, respectively, of the canonical linear mapping $u\mapsto uM$ associated with~$M$.
Moreover, for any subset $S\subseteq\F^\ell$ we denote by $\spann{S}$ the $\F$-linear 
subspace generated by~$S$. 
If $S=\{a_1,\ldots,a_t\}$ is finite we simply write $\spann{a_1,\ldots,a_t}$ for 
$\spann{S}$. 
We will also use the notation $\spann{a, U}:=\spann{a}+U$ for any $a\in \F^\ell$ and any linear subspace $U\subseteq\F^\ell$. 

%%%%%%%%%%%%%%%%%%%%%%%%%%%%%%%%%%%%%%%%%%%%%%%%%%%%%%
\section{Preliminaries}\label{S-Prelim}
\setcounter{equation}{0}
%%%%%%%%%%%%%%%%%%%%%%%%%%%%%%%%%%%%%%%%%%%%
The controller canonical form of an encoder is a well-known means of describing convolutional codes.
Since our paper is completely based on this description we will first present the definition of the controller canonical form and thereafter discuss some of the basic properties as needed later on. 
It also allows us to conveniently introduce the two block codes associated with a convolutional code that are crucial for our investigation.

%%%%%%%%%%%%%%%%%%%%%%%%%%%%%%%%%%%%%%%%%%%%
\begin{definition}\label{D-CCF}
Let $G\in\F[z]^{k\times n}$ be a basic and minimal matrix with Forney indices
$\delta_1,\,\ldots,\delta_r>0=\delta_{r+1}=\ldots=\delta_k$
and degree $\delta:=\sum_{i=1}^k\delta_i$.
Let~$G$ have the rows $g_i=\sum_{\nu=0}^{\delta_i}g_{i,\nu}z^{\nu},\,i=1,\ldots,k,$ where 
$g_{i,\nu}\in\F^n$.
For $i=1,\ldots,r$ define the matrices 
\[
 A_i=\left(\begin{smallmatrix} 0&1& & \\ & &\ddots& \\& & &1\\ & & &0\end{smallmatrix}\right)
      \in\F^{\delta_i\times\delta_i},\ 
 B_i=\begin{pmatrix}1&0&\cdots&0\end{pmatrix}\in\F^{\delta_i},\ 
 C_i=\begin{pmatrix}g_{i,1}\\ \vdots\\ g_{i,\delta_i}\end{pmatrix}\in\F^{\delta_i\times n}.
\]
Then the {\em controller canonical form\/} of~$G$ is defined as 
the matrix quadruple 
$(A,B,C,D)\in\F^{\delta\times\delta}\times\F^{k\times\delta}\times
             \F^{\delta\times n}\times\F^{k\times n}$ 
where 
\[
   A=\left(\begin{smallmatrix} A_1&  & \\ &\ddots &\\ & &A_r\end{smallmatrix}\right),\:
   B=\begin{pmatrix}\bar{B}\\0\end{pmatrix}\text{ with }
   \bar{B}=\left(\begin{smallmatrix} 
            B_1\!\!& &\\ &\ddots & \\ & &\!\!B_r\end{smallmatrix}\right),\:
   C=\left(\begin{smallmatrix}C_1\\ \vdots\\C_r\end{smallmatrix}\right),\:
   D=\left(\begin{smallmatrix}g_{1,0}\\ \vdots\\g_{k,0}\end{smallmatrix}\right)=G(0).
\]
\end{definition}
%%%%%%%%%%%%%%%%%%%%%%%%%%%%%%%%%%%%%%%%%%%
As is made precise next the controller canonical form describes the encoding process of the matrix~$G$ in form of a state space system.
%%%%%%%%%%%%%%%%%%%%%%%%%%%%%%%%%%%%%%%%%%%
\begin{remark}\label{R-CCF}
It is easily seen~\cite[Prop.~2.1, Thm.~2.3]{GL05p} that $G(z)=B(z^{-1}I-A)^{-1}C+D$.
As a consequence, one has for $u=\sum_{t\geq 0}u_tz^t\in\F[z]^k$ and $v=\sum_{t\geq0}v_tz^t\in\F[z]^n$
\[
   v=uG
  \Longleftrightarrow 
  \left\{\begin{array}{rcl} x_{t+1}&=&x_tA+u_tB\\v_t&=&x_tC+u_tD\end{array}
	\;\text{ for all }t\geq0\right\} \text{ where }x_0=0.
\]
\end{remark}
%%%%%%%%%%%%%%%%%%%%%%%%%%%%%%%%%%%%%%%%%%%%

From now on we will assume our data to be as follows.
%%%%%%%%%%%%%%%%%%%%%%%%%%%%%%%%%%%%%%%%%%%%%
\begin{ass}\label{A-dataC}
Let $\cC\subseteq\F[z]^n$ be an $(n,k,\delta)$ code with minimal encoder matrix $G\in\F[z]^{k\times n}$. 
Furthermore, assume that the Forney indices of~$\cC$ are given by 
$\delta_1,\,\ldots,\,\delta_r>0=\delta_{r+1}=\ldots=\delta_k$ and 
let $(A,B,C,D)$ be the corresponding controller canonical form.
\end{ass}
%%%%%%%%%%%%%%%%%%%%%%%%%%%%%%%%%%%%%%%%%%%%%%  
The two index sets 
\begin{equation}\label{e-IJ}
 \cI:=\{1, 1+\delta_1, 1+\delta_1+\delta_2,\ldots , {\TS1+\sum_{i=1}^{r-1}\delta_i}\},\quad 
 \cJ:=\{\delta_1,\delta_1+\delta_2,\ldots, {\TS\sum_{j=1}^r\delta_j=\delta}\}
\end{equation}
will be helpful in the sequel. 
One easily derives the following properties. 
%%%%%%%%%%%%%%%%%%%%%%%%%%%%%%%%%%%%%%%%%%%%%%%
\begin{remark} \label{remarkconcan} \
One has $AB^t=0$ and  
$BB^t=\Smallfourmat{I_r}{0}{0}{0}$. Furthermore, $\im B=\spann{e_i\mid i\in\cI}$ and 
$\ker B=\im(0_{(k-r)\times r},\,I_{k-r})\subseteq\F^k$. Finally, 
\[
   (B^t B)_{i,j}=\begin{cases}1,&\!\text{if } i=j\in \cI\\0, &\!\text{else,}\end{cases},
   \
   (A^t A)_{i,j}=\begin{cases}1, &\!\text{if } i=j\notin \cI \\ 0, &\!\text{else,}\end{cases},
    \
   (A A^t)_{i,j}=\begin{cases}1, &\!\text{if } i=j\notin \cJ \\ 0, &\!\text{else.}\end{cases}   
\]
As a consequence, $A^tA+B^tB=I_{\delta}$.
\end{remark}
%%%%%%%%%%%%%%%%%%%%%%%%%%%%%%%%%%%%%

The following two block codes will play a crucial role throughout the paper.
%%%%%%%%%%%%%%%%%%%%%%%%%%%%%%%%%%%%
\begin{definition}\label{D-Cconst}
For~$\cC$ as above we define
$\cCconst:=\cC\cap\F^n$ to be the block code consisting of the constant codewords in~$\cC$.
Moreover, let $\CC:=\im\Smalltwomat{C}{D}\subseteq\F^n$ and define $\widehat{r}\in\{0,\ldots,n-k\}$ such that $\dim\CC=k+\widehat{r}$.
\end{definition}
%%%%%%%%%%%%%%%%%%%%%%%%%%%%%%%%%%%%
The following properties of these codes are easily seen from the controller canonical form.

%%%%%%%%%%%%%%%%%%%%%%%%%%%%%%%%%%
\begin{remark}\label{R-CC}
\begin{arabiclist}
\item Suppose the encoder matrix $G$ is as in Definition~\ref{D-CCF}. Then 
	$\CC=\im\Smalltwomat{C}{D}=\spann{g_{i,\nu}\mid i=1,\ldots,k,\,\nu=0,\ldots,\delta_i}$.
        Recalling that two different encoders of~$\cC$ differ only by a left unimodular
	transformation it follows immediately that the block code~$\CC$ 
        does not depend on the choice of the encoder~$G$ but rather is an invariant 
	of the code~$\cC$. 
	Since $\rank D=k$ it is clear that the dimension of~$\CC$ is indeed at least~$k$. 
\item One has $\dim\cCconst=k-r$ and precisely, with the notation from~(1), 
	\begin{equation}\label{e-V}
	  \cCconst=\spann{g_i\mid i=r+1,\ldots,k}=(\ker B)D:=\{uD\mid u\in\ker B\}.
	\end{equation}
	This also shows $\cCconst\subseteq\CC$.
	Furthermore we have $\im D=\im B^tD\oplus\cCconst$.
\end{arabiclist}
\end{remark}

%%%%%%%%%%%%%%%%%%%%%%%%%%%%%%%%%%%%

In accordance with block code theory the dual of a convolutional code is defined with respect to the canonical bilinear form
\[
  \beta:\,\F[z]^n\times\F[z]^n\longrightarrow\F[z],\quad
  \big((a_1,\ldots,a_n),(b_1,\ldots,b_n)\big)\longmapsto\sum_{j=1}^n a_jb_j.
\]
With this notation the {\em dual code\/} is explicitly defined as
\begin{equation}\label{e-Cperpdef}
  \widehat\cC:=\{w\in\F[z]^n\mid \beta(w,v)=0\text{ for all }v\in\cC\}.
\end{equation}
In the sequel we will also let~$\beta$ denote the canonical bilinear form on $\F^{\ell}$ for any $\ell\in\N$.
In that case we will use the notation 
$U^{\perp}:=\{v\in\F^\ell\mid \beta(v,u)=0\text{ for all } u\in U\}\subseteq\F^\ell$ 
for the orthogonal of a subspace $U\subseteq\F^\ell$. 
The different notation $\widehat{\cC}$ versus $U^{\perp}$ for the dual of a convolutional code~$\cC$ versus a block code~$U$ is simply to avoid cumbersome notation later on.
It is well known \cite[Thm.~7.1]{McE98a} that 
\begin{equation}\label{e-Cperp}
 \text{if $\cC$ is an $(n,k,\delta)$ code, then $\widehat\cC$ is an 
       $(n,n-k,\delta)$ code.}
\end{equation} 
The two block codes from Definition~\ref{D-Cconst} and the corresponding objects $\CCperp$ and $\cCperpconst$ for the dual code~$\widehat{\cC}$ behave as follows under duality. 
%%%%%%%%%%%%%%%%%%%%%%%%%%%%%%
\begin{proposition} \label{duality} 
One has $(\CC)^\bot=\cCperpconst$.
As a consequence, $\widehat\cC$ has exactly $n-k-\widehat{r}$ zero Forney indices and 
$\widehat{r}$ nonzero Forney indices. 
Moreover, $\dim\CCperp=n-k+r$.
\end{proposition}
%%%%%%%%%%%%%%%%%%%%%%%%%%%%%%%%
\begin{proof} 
Using the notation and statement of Remark~\ref{R-CC}(1) we obtain
\begin{align*}
  c\in(\CC)^\bot&\Longleftrightarrow \beta(c,g_{i,\nu})=0
             \text{ for all }i=1,\ldots,k,\,\nu=0,\ldots,\delta_i\\
	     &\Longleftrightarrow\beta(c,g_i)=0\text{ for all }i=1,\ldots,k
	     \Longleftrightarrow c\in\widehat{\cC}\cap\F^n
	     =\cCperpconst,
\end{align*}
where the second equivalence uses the fact that~$c$ is a constant vector. 
The consequences are clear from the definition of~$r$ and~$\widehat{r}$.
\end{proof}

%%%%%%%%%%%%%%%%%%%%%%%%%%%%%%%%%%%%%%%%%
\begin{exa}\label{example1} 
Let $q=2$, $n=5$, $k=2$, and $\cC\subseteq\F_2[z]^5$ be the code generated by the basic and minimal encoder 
$G=\begin{pmatrix}1+z+z^3&z^2&z^2&1&z\\1&1&0&1&0\end{pmatrix}$.
Thus $\delta=3,\,\delta_1=3,\,\delta_2=0$, and $r=1$.
The associated controller canonical form is 
\[
   A=\begin{pmatrix}0&1&0\\0&0&1 \\0&0&0\end{pmatrix},\; 
   B=\begin{pmatrix}1&0&0\\ 0&0&0\end{pmatrix},\; C=\begin{pmatrix}1&0&0&0&1\\0&1&1&0&0\\1&0&0&0&0\end{pmatrix},\; D=\begin{pmatrix}1&0&0&1&0\\1&1&0&1&0\end{pmatrix}.
\]
Using Equation\eqnref{e-V} we obtain $\cCconst=\im\begin{pmatrix}1&1&0&1&0\end{pmatrix}$ while $\CC=\F_2^5$. 
As a consequence, $\widehat r=3$. 
It can easily be checked that the dual code $\widehat\cC$ is generated by the basic and minimal matrix
$\widehat G=\begin{pmatrix} 1&z&0&1+z&0 \\ 0&z&z&z&1 \\  0&0&1&0&z\end{pmatrix}$.
Indeed, it has $\widehat{r}=3$ nonzero Forney indices as stated in Proposition~\ref{duality}.
The controller canonical form is 
\[
  \widehat A=0,\; 
  \widehat B=I_3,\;
  \widehat C=\begin{pmatrix} 0&1&0&1&0\\0&1&1&1&0 \\0&0&0&0&1 \end{pmatrix},\; 
  \widehat D=\begin{pmatrix} 1&0&0&1&0\\0&0&0&0&1 \\0&0&1&0&0 \end{pmatrix}.
\]
Moreover,  
\[
   \CCperp=\im\begin{pmatrix}0&0&1&0&0\\0&0&0&0&1 \\0&1&0&1&0 \\ 1&0&0&1&0\end{pmatrix},\
   \cCperpconst=\{0\}.
\]
This is indeed in compliance with Proposition~\ref{duality} since
$(\CC)^{\perp}=\cCperpconst$ and $(\CCperp)^{\perp}=\cCconst$.
\end{exa}
%%%%%%%%%%%%%%%%%%%%%%%%%%%%%%%%%%%%%%%%%%

Let us now return to the general case. 
Block code theory allows us to apply the MacWilliams transformation to the block codes in
Proposition~\ref{duality}. 
Before doing so it will be useful to define the weight enumerator for arbitrary affine 
sets in $\F^n$ as it will be needed in the following sections.
Recall the Hamming weight $\wt(a)$ for $a\in\F^n$. 

%%%%%%%%%%%%%%%%%%%%%%%%%%%%%%%%%%%%%%%%%
\begin{definition}\label{D-we}
Let $\C[W]_{\leq n}$ denote the vector space of polynomials over~$\C$ of degree at most~$n$.
For any affine subspace $S\subseteq\F^n$ we define the {\em weight enumerator\/} of~$S$
to be the polynomial 
$\we(S):=\sum_{j=0}^n \alpha_jW^j\in\C[W]_{\leq n}$, where $\alpha_j:=\#\{a\in S\mid \wt(a)=j\}$. We also put $\we(\emptyset)=0$.
\end{definition}
%%%%%%%%%%%%%%%%%%%%%%%%%%%%%%%%%%%%%%%%%%
Recall that the classical MacWilliams Identity for block codes states that 
for $k$-dimensional codes $\cC\subseteq\F^n=\F_q^n$ one has
\begin{equation}\label{e-MacWBC}
   \we(\cC^{\perp})=q^{-k}\HH\big(\we(\cC)\big),
\end{equation}
where $\HH$ is the MacWilliams transformation
\begin{equation}\label{e-h}
  \HH:\, \C[W]_{\leq n}\longrightarrow \C[W]_{\leq n},\quad
       \HH(f)(W):=(1+(q-1)W)^n f\big({\TS\frac{1-W}{1+(q-1)W}}\big).    
\end{equation} 
Observe that the mapping~$\HH$ is $\C$-linear and satisfies $\HH^2(f)=q^nf$.
It should be kept in mind that~$\HH$ depends on the parameters~$n$ and~$q$.
Since throughout this paper these parameters will be fixed we do not indicate them explicitly.

Let us now return to convolutional codes. 
Using\eqnref{e-MacWBC} and Proposition~\ref{duality} one immediately obtains 
%%%%%%%%%%%%%%%%%%%%%%%%%%%%%%%%%%%%%%%%%%
\begin{corollary}\label{C-MacWCC}
$q^{k+\hat{r}}\we(\cCperpconst)=\HH\big(\we(\CC)\big)$. 
\end{corollary}
%%%%%%%%%%%%%%%%%%%%%%%%%%%%%%%%%%%%%%%%%%%

%%%%%%%%%%%%%%%%%%%%%%%%%%%%%%%%%%%%%%%%%%%%
\section{The Adjacency Matrix of a Code}\label{S-adjmatrix}
\setcounter{equation}{0}
%%%%%%%%%%%%%%%%%%%%%%%%%%%%%%%%%%%%%%%%%%%%%%%%%%%%%%%
The (weight) adjacency matrix as defined next has been introduced in \cite{McE98a} and 
studied in detail in \cite{GL05p}.
The aim of this section is to survey the structure and redundancies of the adjacency 
matrix for a given convolutional code.
Let the data be as in\eqnref{e-Fq} and General Assumption~\ref{A-dataC}. 
Recall from Remark~\ref{R-CCF} that the controller canonical form leads to a state space description of the encoding process where the input is given by the coefficients of the 
message stream while the output is the sequence of codeword coefficients. 
The following matrix collects for each possible pair of states~$(X,Y)$ the information whether via a suitable input~$u$ a transition from~$X$ to~$Y$ is possible, i.~e., whether $Y=XA+uB$ for some~$u$, and if so, collects the weights of all associated outputs $v=XC+uD$.

%%%%%%%%%%%%%%%%%%%%%%%%%%%%%
\begin{definition}\label{Deflambda} 
We call~$\F^{\delta}$ the {\em state space\/} of the encoder~$G$ (or of the controller canonical form). 
Define $\cF:=\F^\delta\times\F^\delta$. 
The {\em (weight) adjacency matrix\/} 
$\Lambda(G)=(\lambda_{X,Y})\in\C[W]^{q^{\delta}\times q^{\delta}}$ is 
defined to be the matrix indexed by $(X,Y)\in\cF$ with the entries 
\[
     \lambda_{X,Y}:=\we(\{XC+uD\mid u\in\F^k: Y=XA+uB\})\in\C[W]_{\leq n}.
\]
A pair of states $(X,Y)\in\mathcal F$ is called {\em connected\/} if $\lambda_{X,Y}\neq 0$, else it is called {\em disconnected}. 
The set of all connected state pairs is denoted by $\Delta\subseteq\cF$. 
\end{definition}
%%%%%%%%%%%%%%%%%%%%%%%%%%%%%%%%%
Observe that in the case $\delta=0$ the matrices $A,\,B,\,C$ do not exist while $D=G$. 
As a consequence, $\Lambda=\lambda_{0,0}=\we(\cC)$ is the ordinary weight enumerator of the block code $\cC=\{uG\mid u\in\F^k\}\subseteq\F^n$.

%%%%%%%%%%%%%%%%%%%%%%%%%%%%%%%%%%%%%%%
\begin{exa}\label{E-exa2}
Let the data be as in Example~\ref{example1}. 
In order to explicitly display the adjacency matrices corresponding to~$G$ and~$\widehat{G}$ we need to fix an ordering on the state space~$\F_2^3$. 
Let us choose the lexicographic ordering, that is, we will order the row and column indices according to 
\begin{equation}\label{e-lex}
  (0,0,0),\,(0,0,1),\,(0,1,0),\,(0,1,1),\,(1,0,0),\,(1,0,1),\,(1,1,0),\,(1,1,1).
\end{equation}
Then it is lengthy, but straightforward to see that 
\[
   \Lambda(G)=
     \left( \begin{smallmatrix}1+W^3&0&0&0&W+W^2&0&0&0 \\W+W^2&0&0&0&W+W^2&0&0&0 \\0&W^2+W^3&0&0&0&W+W^4&0&0 \\0&W^2+W^3&0&0&0&W^2+W^3&0&0 \\0&0&W^2+W^3&0&0&0&W^2+W^3&0 \\0&0&W+W^4&0&0&0&W^2+W^3&0 \\0&0&0&W^3+W^4&0&0&0&W^3+W^4 \\0&0&0&W^3+W^4&0&0&0&W^2+W^5 \end{smallmatrix}\right).
\]
For instance, in order to compute the entry in the $4$th row and $2$nd column put 
$X:=(0,1,1),\,Y:=(0,0,1)$. 
Using the controller canonical form as given in Example~\ref{example1} one has $XA+uB=Y$ if and only if $u\in\{(0,0),\,(0,1)\}$ and thus 
\[
  \lambda_{X,Y}
         =\we\big\{XC+uD\,\big|\,u\in\{(0,0),\,(0,1)\} \big\}
	 =\we\{(1,1,1,0,0),\,(0,0,1,1,0)\}=W^2+W^3.
\]
Likewise we obtain for the dual code 
\[
  \Lambda(\widehat G)=
  \begin{pmatrix}1&W&W&W^2&W^2&W^3&W^3&W^4\\W&W^2&1&W&W^3&W^4&W^2&W^3\\
     W^3&W^2&W^4&W^3&W^3&W^2&W^4&W^3\\W^4&W^3&W^3&W^2&W^4&W^3&W^3&W^2 \\
     W^2&W^3&W^3&W^4&W^2&W^3&W^3&W^4 \\W^3&W^4&W^2&W^3&W^3&W^4&W^2&W^3 \\
     W&1&W^2&W&W^3&W^2&W^4&W^3 \\W^2&W&W&1&W^4&W^3&W^3&W^2 
   \end{pmatrix}.
\]
Later in Theorem~\ref{T-MacWDuality} we will see that these two adjacency matrices determine each other in form of a generalized MacWilliams identity.
\end{exa}
%%%%%%%%%%%%%%%%%%%%%%%%%%%%%%%%%%%%%%%

%%%%%%%%%%%%%%%%%%%%%%%%%%%%%%%%%%%%%%%
\begin{remark}\label{R-weightdistr}
The adjacency matrix contains very detailed information about the code. 
Firstly, it is well-known that the classical path weight enumerator of a convolutional code
\cite[p.~154]{JoZi99}
can be computed from the adjacency matrix, for details see in~\cite{McE98a},~\cite[Thm.~3.8]{GL05p}, and \cite[Sec.~3.10]{JoZi99}.
Secondly, at the end of Section~3 in~\cite{GL05p} it has been outlined that 
the extended row distances~\cite{JPB90} as well as the active burst distances~\cite{HJZ02} can be recovered from the adjacency matrix. 
As explained in~\cite{JPB90},~\cite{HJZ02} these parameters are closely related to the error-correcting performance of the code.
\end{remark}
%%%%%%%%%%%%%%%%%%%%%%%%%%%%%%%%%%%%%%%%

It is clear from Definition~\ref{Deflambda} that the adjacency matrix depends on the chosen 
encoder~$G$. This dependence, however, can nicely be described.
Since we will make intensive use of the notation later on we introduce the following.
%%%%%%%%%%%%%%%%%%%%%%%%%%%%%%%%%%%%%%%%%%%%
\begin{definition}\label{D-PP}
For any $P\in \text{GL}_{\delta}(\F)$ define $\cP(P)\in \text{GL}_{q^{\delta}}(\C)$ by 
$\cP(P)_{X,Y}=1$ if $Y=XP$ and $\cP(P)_{X,Y}=0$ else.
Furthermore, let $\Pi:=\{\cP(P)\mid P\in \text{GL}_{\delta}(\F)\}$ denote the subgroup of all such permutation matrices. 
\end{definition}
%%%%%%%%%%%%%%%%%%%%%%%%%%%%%%%%%%%%%%%%
By definition, the matrix $\cP(P)$ corresponds to the permutation on the set $\F^{\delta}$
induced by the isomorphism~$P$.
Notice that~$\cP$ is an isomorphism of groups and basically is the canonical faithful permutation representation of the group $\text{GL}_{\delta}(\F)$.
Obviously, we have for any $\Lambda\in\C[W]^{q^{\delta}\times q^{\delta}}$ and any 
$\cP:=\cP(P)\in\Pi$ the identity
\begin{equation}\label{e-Pitrafo}
     \big(\cP\Lambda\cP^{-1}\big)_{X,Y}=\Lambda_{XP,YP}\text{ for all }(X,Y)\in\cF.
\end{equation}
Now we can collect the following facts about the adjacency matrix. 
%%%%%%%%%%%%%%%%%%%%%%%%%%%%%%%%%%%%%%%
\begin{remark} \label{homogeneous}
\
\begin{alphalist}
\item Using the obvious fact $\wt(\alpha v)=\wt(v)$ for any $\alpha\in\F^*$ and $v\in\F^n$ one 
      immediately has 
      $\lambda_{X,Y}=\lambda_{\alpha X,\alpha Y}$ for all $\alpha\in\F^*$. 
      Hence $\Lambda(G)$ is invariant under conjugation with permutation matrices that are induced by scalar multiplication on $\F^{\delta}$, i.~e., under conjugation with matrices $\cP(P)$ where 
      $P=\alpha I$ for some $\alpha\in\F^*$.
\item In \cite[Thm.~4.1]{GL05p} it has been shown that if $G_1,\,G_2\in\F[z]^{k\times n}$ 
      are two minimal encoders of~$\cC$ then 
      $\Lambda(G_1)=\cP\Lambda(G_2)\cP^{-1}$ for some $\cP\in\Pi$. 
      Hence the equivalence class of $\Lambda(G)$ modulo conjugation by~$\Pi$, where~$G$ is any minimal encoder, forms an invariant of the code. It is called the 
      {\em generalized adjacency matrix\/} of~$\cC$.
\item Combining~(b) and~(a) we see that the equivalence class of $\Lambda(G)$ is already fully obtained by conjugating $\Lambda(G)$ with matrices $\cP(P)$ where $P$ is in the projective linear group 
      $\text{GL}_{\delta}(\F)/\{\alpha I\mid \alpha\in\F^*\}$. 
      This reduces the computational effort when computing examples.
\end{alphalist}
\end{remark}
%%%%%%%%%%%%%%%%%%%%%%%%%%%%%%%%%%%%%%%%

Let us now return to Definition~\ref{Deflambda}. 
Notice that $(X,Y)\in\cF$ is connected if and only if there exists some $u\in\F^k$ such that $(X,Y)=(X,XA+uB)$. 
Using $\rank B=r$ we obtain 
%%%%%%%%%%%%%%%%%%%%%%%%%%%%%%%%%%%
\begin{proposition}\label{P-Delta} 
$\Delta=\im \Smallfourmat{I}{A}{0}{B}$ is an $\F$-vector space of dimension $\delta +r$. 
\end{proposition}
%%%%%%%%%%%%%%%%%%%%%%%%%%%%%%%%%%%
Later on we will also need the dual of $\Delta$ in $\cF$. 
With the help of Remark~\ref{remarkconcan} it can easily be calculated and is given as follows.
%%%%%%%%%%%%%%%%%%%%%%%%%%%%%%%%%%
\begin{lemma} \label{Deltabot}
$\Delta^\bot=\{(X,-XA)\mid X=(X_1,\ldots,X_{\delta})\in \F^\delta\text{ such that } X_j=0
 \text{ for } j\in \cJ\}$. 
\end{lemma}
%%%%%%%%%%%%%%%%%%%%%%%%%%%%%%%%%%%%%%

In the next lemma will show that the nontrivial entries~$\lambda_{X,Y}$ of the adjacency matrix can be described as weight enumerators of certain cosets of the block code~$\cCconst$. 
More precisely, we will relate them to the $\F$-vector space homomorphism 
\begin{equation}\label{e-phi}
  \varphi:\cF\longrightarrow \F^n, \quad(X,Y)\longmapsto XC+YB^tD.
\end{equation}
Recall the notation $\spann{a,U}$ as introduced at the end of Section~\ref{S-Intro}. 
%%%%%%%%%%%%%%%%%%%%%%%%%%%%%%%%%%%
\begin{lemma} \label{weightenum} 
For any state pair $(X,Y)\in \Delta$ we have 
$\lambda_{X,Y}=\we\big(\varphi(X,Y)+\cCconst\big)$.
Moreover, 
\[
   \lambda_{X,Y}=\begin{cases}
       \we(\cCconst), &\text{if } \varphi(X,Y)\in \mathcal \cCconst, \\[1ex] 
       \frac 1{q-1}\Big(\we\big(\spann{\varphi(X,Y),\,\cCconst}\big)-\we(\cCconst)\Big),& \text{else.}\end{cases}
\]
\end{lemma}
%%%%%%%%%%%%%%%%%%%%%%%%%%%%%%%%%%
\begin{proof}
First notice that for any $(X,Y)\in\Delta$ the set $\{u\in\F^k\mid Y-XA=uB\}$ is non-empty.
Right-multiplying the defining equation of this set with $B^t$ we get upon use of
Remark~\ref{remarkconcan} that $YB^t=uBB^t$, which says that the first~$r$ entries of~$u$ are completely determined by~$Y$. 
This shows $\{u\in\F^k\mid Y-XA=uB\}\subseteq YB^t+\im(0,\,I_{k-r})$.
From $\im(0,\,I_{k-r})=\ker B$, see Remark~\ref{remarkconcan}, we conclude that these two affine subspaces coincide. 
Hence, using Remark~\ref{R-CC}(2), we obtain
\[
  \lambda_{X,Y}=\we(XC+(YB^t+\ker B)D)=\we\big(\varphi(X,Y)+(\ker B)D\big)=
  \we(\varphi(X,Y)+\cCconst).
\] 
This shows the first part of the lemma. 
If $\varphi(X,Y)\in \cCconst$, we immediately conclude $\lambda_{X,Y}=\we(\cCconst)$. 
Otherwise we have $\lambda_{X,Y}=\we(\varphi(X,Y)+\cCconst)=\we\big(\alpha(\varphi(X,Y)+\cCconst)\big)
=\we\big(\alpha\varphi(X,Y)+\cCconst\big)$ for all $\alpha\in\F^*$. 
Moreover,  
\[
    \spann{\varphi(X,Y),\,\cCconst}=
    \bigcup_{\alpha\in\F}\big(\alpha\varphi(X,Y)+\cCconst\big),
\]
where due to $\varphi(X,Y)\notin\cCconst$ this union is disjoint. 
From this the last assertion can be deduced.
\end{proof}

The lemma shows that the mapping~$\varphi$ and the block code $\cCconst$ 
along with the knowledge of~$\Delta$ fully determine $\Lambda(G)$.
Moreover, to find out how many state pairs $(X,Y)\in\Delta$ are mapped to $\cCconst$, we 
will slightly modify the mapping $\varphi$.
%%%%%%%%%%%%%%%%%%%%%%%%%%%%%%
\begin{lemma} \label{hom}
The homomorphism \[
   \Phi:\Delta\longrightarrow \CC/\cCconst,\quad (X,Y) 
   \longmapsto \varphi(X,Y)+\cCconst
\]
is well-defined, surjective and satisfies
\begin{alphalist}
\item $\ker\Phi=\{(X,Y)\in\Delta\mid \varphi(X,Y)\in\cCconst\}$,
\item $\dim\ker\Phi=\delta-\widehat{r}$, where~$\widehat{r}$ is as in 
      Definition~\ref{D-Cconst}.
\end{alphalist}
\end{lemma}
%%%%%%%%%%%%%%%%%%%%%%%%%%%%%%%%%%%%%%%%%%%%%%%%
\begin{proof}
The well-definedness of $\Phi$ simply follows from $\im\varphi\subseteq \CC$. 
As for the surjectivity notice that any row of $\Smalltwomat{C}{D}$ that is not in $\cCconst$
is a row of the matrix $\Smalltwomat{C}{BB^tD}$, see also Remark~\ref{R-CC}(2). 
Moreover, by Remark~\ref{remarkconcan} we have $\im\Smalltwomat{C}{BB^tD}=\im\Smallfourmat{I}{A}{0}{B}\Smalltwomat{C}{B^tD}=\varphi(\Delta)$, where the latter follows from the definition of the mapping $\varphi$ along with Proposition~\ref{P-Delta}.
All this implies the surjectivity of~$\Phi$. 
Now part~(a) is trivial.
The surjectivity together with $\dim\Delta=\delta+r$ yields~(b) since 
$\dim\CC=k+\widehat{r}$ while $\dim\cCconst=k-r$. 
\end{proof}

Let us illustrate the results so far by the previous example.
%%%%%%%%%%%%%%%%%%%%%%%%%%%%%%%%%%%%%%%%%%%%%%%%%%%%%%%%%%%%%
\begin{exa}\label{E-Exadata}
Consider again the data from Example~\ref{example1} and~\ref{E-exa2}.
We can observe the following properties of the two adjacency matrices.
\begin{arabiclist}
\item The matrix~$\Lambda(G)$ has exactly $2^4=16$ nonzero entries, while   
      $\Lambda(\widehat{G})$ has exactly $2^6=64$ nonzero entries. 
      This is in compliance with Proposition~\ref{P-Delta} applied to~$\cC$ as well 
      as $\widehat{\cC}$.
\item Each nonzero entry of~$\Lambda(G)$ is the sum of two monomials, while each entry of
      $\Lambda(\widehat{G})$ is a monomial. 
      This also follows from the first part of Lemma~\ref{weightenum} since 
      $\#\cCconst=2$ while $\cCperpconst=\{0\}$. 
\item There are~$4$ entries in~$\Lambda(\widehat{G})$ that are equal to~$1$. 
      This also follows from application of Lemma~\ref{hom} and Lemma~\ref{weightenum}
      to the dual code: we obtain $2^{\delta-r}=4$ times the case $\widehat{\lambda}_{X,Y}=\we(\cCperpconst)=1$ while the second case appearing in 
      Lemma~\ref{weightenum}, being the difference of the weight enumerators of two block codes, never contains the monomial $1=W^0$. 
      Along the same line of arguments one can also explain that~$\lambda_{0,0}$ is the only entry of~$\Lambda(G)$ containing the monomial~$1=W^0$. 
\end{arabiclist}
\end{exa}
%%%%%%%%%%%%%%%%%%%%%%%%%%%%%%%%%%%%%%%

As a consequence of Lemma~\ref{hom} one has 
\begin{equation}\label{e-imphi}
  \varphi(\Delta)+\cCconst=\CC.
\end{equation}
We are now prepared to clarify some more redundancies in the adjacency matrix of~$\cC$.
%%%%%%%%%%%%%%%%%%%%%%%%%%%%%%%%
\begin{proposition}\label{linearity}
Let $\Delta^*\subseteq\Delta$ be any subspace such that $\Delta=\Delta^*\oplus\ker\Phi$.
Moreover, define $\Delta^-:=\spann{(0,e_i)\mid i\notin\cI}\subseteq\cF$.  
Then 
\begin{alphalist}
\item $\Delta\oplus\Delta^-=\cF$, hence $\Delta^*\oplus\ker\Phi\oplus\Delta^-=\cF$.
\item For $(X,Y)\in\Delta^-$ and $(X',Y')\in\Delta$ one has 
      $\lambda_{X+X',Y+Y'}=0$ if and only if $(X,Y)\not=0$.
\item For $(X,Y)\in\Delta^*$ and $(X',Y')\in\ker\Phi$ one has 
      $\lambda_{X+X',Y+Y'}=\lambda_{X,Y}$.
\end{alphalist}
\end{proposition}
%%%%%%%%%%%%%%%%%%%%%%%%%%%%%%%%%%%%%%%%%
\begin{proof} 
(a) $\Delta\cap\Delta^-=\{0\}$ follows from $e_i\notin\im B$ for $i\notin\cI$. The rest is clear since $\dim\Delta^-=\delta-r=2\delta-\dim\Delta$.
(b) is obvious from the first direct sum in~(a) and the definition of~$\Delta$.
As for~(c) notice that by linearity and Lemma~\ref{hom}(a) $\varphi(X,Y)-\varphi(X+X',Y+Y')\in\cCconst$. 
Hence $\varphi(X,Y)+\cCconst=\varphi(X+X',Y+Y')+\cCconst$ and the result follows from Lemma~\ref{weightenum}.
\end{proof}

Concerning Proposition~\ref{linearity}(c) it is worth mentioning that the converse statement 
$[\lambda_{X,Y}=\lambda_{\tilde{X},\tilde{Y}} \Longrightarrow (X,Y)-(\tilde{X},\tilde{Y})\in\ker\Phi]$ is in general not true as different affine sets may well have the same weight enumerator. 
Moreover, notice that the results above are obviously true for any direct complement 
of $\Delta$ in $\cF$.
Our particular choice of~$\Delta^-$ will play an important role due to the following 
corollary.

%%%%%%%%%%%%%%%%%%%%%%%%%%%%%
\begin{corollary}\label{C-Deltastar}
One has $\varphi|_{\Delta^-}=0$ and 
$\CC=\bigcup_{(X,Y)\in\Delta^*}\big(\varphi(X,Y)+\cCconst)$ with the union being disjoint.
\end{corollary}
%%%%%%%%%%%%%%%%%%%%%%%%%%%%%%%%%%%%%
\begin{proof}
The first part follows directly from the definition of all objects involved. 
The inclusion ``$\supseteq$'' of the second statement is obvious. 
For the other inclusion let $XC+uD\in\CC$ for some $(X,u)\in\F^{\delta+k}$. 
Using that $\im D=\im B^tD+\cCconst$, see Remark~\ref{R-CC}(2), this yields 
$XC+uD=XC+YB^tD+a$ for some $Y\in\F^{\delta}$ and $a\in\cCconst$. Hence 
$XC+uD\in\varphi(X,Y)+\cCconst$ where $(X,Y)\in\cF$. 
Now $\varphi|_{\Delta^-}=0$ and Lemma~\ref{hom}(a) imply that without loss of generality $(X,Y)\in\Delta^*$. 
The disjointness of the union follows from $\Delta^*\cap\ker\Phi=\{0\}$ 
with the same lemma.
\end{proof}

We will conclude this chapter by computing the sum over all entries of the adjacency matrix of a convolutional code in order to demonstrate how the terminology developed above facilitates this task.
The result will be needed later on for proving Theorem~\ref{entriesl}.
%%%%%%%%%%%%%%%%%%%%%%%%%%%%%%%%
\begin{proposition} \label{sumoverlambda} 
The entries of the adjacency matrix satisfy
\[\sum_{(X,Y)\in\Delta^*} \lambda_{X,Y}=\we(C_{\mathcal C})\quad \text{and}\quad
  \sum_{(X,Y)\in \cF}\lambda_{X,Y}=\sum_{(X,Y)\in\Delta}\lambda_{X,Y}
     =q^{\delta-\widehat r}\we(C_{\mathcal C}).
\]
\end{proposition}
%%%%%%%%%%%%%%%%%%%%%%%%%%%%%%%%%%%
{\sc Proof:}
Using Lemma~\ref{weightenum}  and Corollary~\ref{C-Deltastar} we obtain
\[
   \sum_{(X,Y)\in\Delta^*} \lambda_{X,Y}=\sum_{(X,Y)\in\Delta^*}\we(\varphi(X,Y)+\cCconst)
   =\we(\CC). 
\]
Next notice that $\sum_{(X,Y)\in \mathcal F}\lambda_{X,Y}=\sum_{(X,Y)\in \Delta}\lambda_{X,Y}$ as any disconnected state pair $(X,Y)$ satisfies $\lambda_{X,Y}=0$. 
Hence with Proposition~\ref{linearity}(c) and Lemma~\ref{hom}(b) we get
\begin{align*}
   \sum_{(X,Y)\in\Delta} \lambda_{X,Y}&
       =\sum_{(\bar X,\bar Y)\in\ker \Phi}\
          \sum_{(X,Y)\in \Delta^*}\lambda_{X+\bar X, Y+\bar Y}
       =\sum_{(\bar X,\bar Y)\in\ker \Phi}\
           \sum_{(X,Y)\in \Delta^*}\lambda_{X,Y}\\
      &=\sum_{(\bar X,\bar Y)\in\ker \Phi}\we(\CC)
       =q^{\delta-\widehat r}\we(\CC).
\end{align*}
\mbox{}\vspace*{-10.5ex}

\hfill$\Box$

\vspace*{3ex}

\noindent It is straightforward to verify the second result of this proposition for Example~\ref{example1}/\ref{E-exa2}.

%%%%%%%%%%%%%%%%%%%%%%%%%%%%%%%%%%%%%%%%%%%%%%%%%%%%%%%
\section{The MacWilliams Matrices}\label{S-MacWMat}
\setcounter{equation}{0}
%%%%%%%%%%%%%%%%%%%%%%%%%%%%%%%%%%%%%%%%%%%%%%%%%%%%%%%
Recall the notation from\eqnref{e-Fq} and fix some $\delta\in\N$. 
In this section we will define a set of complex matrices that are essential for our transformation formula as discussed in the next section, and we will collect some of their properties. 
To define the matrices we will use complex-valued characters on $\F^\delta$, i.~e., group homomorphisms $(\F^\delta,+)\longrightarrow(\C^*,\cdot)$. 
It is a well known fact~\cite[Thm.~5.5]{LiNi97}, that, using a fixed primitive $p$-th root of unity $\zeta\in\C$, the character group on $\F^\delta$ is given as 
$\{\zeta^{\tau(\beta (X,\,\cdot\,))}\mid X\in \F^\delta\}$, where 
$\tau:\F\longrightarrow \F_p$, $a\longmapsto\sum_{i=0}^{s-1}a^{p^i}$
is the usual trace form and~$\beta$ is the canonical bilinear form on~$\F^{\delta}$. 
It will be convenient to define 
\begin{equation}\label{e-thetachar}
  \theta_X:=\zeta^{\tau(\beta(X,\,\cdot\,))}:\F^{\delta}\longrightarrow\C^*
  \text{ for all }X\in\F^{\delta}.
\end{equation}
For easier reference we list the following properties. 
%%%%%%%%%%%%%%%%%%%%%%%%%%%%%%%%%%
\begin{remark}\label{R-thetachar}
\begin{alphalist}
\item The character $\theta_X$ is nontrivial if and only if $X\not=0$. 
	This follows from the fact that for $X\not=0$ we have %$\dim\ker\beta(X,\cdot)=\delta-1$ and hence 
	$\#\im\beta(X,\cdot)=q$, while the $\F_p$-linear and surjective mapping~$\tau$ 
	satisfies $\#\ker\tau=p^{s-1}=\frac{q}{p}$.
\item Applying a standard result on characters~\cite[Thm.~5.4]{LiNi97}
      we have $\sum_{Y\in\F^{\delta}}\theta_X(Y)=0$ 
      if $X\not=0$ while $\sum_{Y\in\F^{\delta}}\theta_0(Y)=q^{\delta}$. 
\item For all $X,\,Y\in\F^{\delta}$ and all $P\in \text{GL}_{\delta}(\F)$ we have 
      $\theta_X(Y)=\theta_Y(X)$ and 
      $\theta_{XP}(Y)=\theta_X(YP^t)$.
\item For all $X,\,Y,\,Z_1,\,Z_2\in\F^{\delta}$ one has 
      $\theta_X(Z_1)\theta_Y(Z_2)=\theta_{(X,Y)}(Z_1,Z_2)$ where the latter is defined on 
      $\F^{2\delta}$ analogously to\eqnref{e-thetachar}, that is,
      $\theta_{(X,Y)}(Z_1,Z_2):=\zeta^{\tau(\beta((X,Y),(Z_1,Z_2)))}$ with~$\beta$ also denoting the canonical bilinear form on~$\F^{2\delta}$.
\end{alphalist}
\end{remark}
%%%%%%%%%%%%%%%%%%%%%%%%%%%%%%%%%%%

%%%%%%%%%%%%%%%%%%%%%%%%%%%%%%%%%%%%%
\begin{definition}\label{D-MacWMat}
Let $\zeta\in\C^*$ be a fixed primitive $p$-th root of unity.
For $P\in \text{GL}_\delta(\F)$ we define the $P$-{\em MacWilliams matrix\/} as
\[
   \cH(P):=q^{-\frac\delta 2}\big(\theta_{XP}(Y)\big)_{(X,Y)\in\cF}
   \in\C^{q^{\delta}\times q^{\delta}}.
\]
For simplicity we also put $\cH:=\cH(I)$. 
For $\delta=0$ we simply have $\cH=1$.
\end{definition}
%%%%%%%%%%%%%%%%%%%%%%%%%%%%%

%%%%%%%%%%%%%%%%%%%%%%%%%%%%%%%%%%%%%%
\begin{remark}\label{R-dependence}
Notice that the MacWilliams matrices depend on~$\delta$. 
Since this parameter will be fixed throughout our paper (except for the examples and Remark~\ref{R-MacWMat}) we will not explicitly denote this dependence. 
Moreover, the matrices depend on the choice of the primitive root~$\zeta$. 
This dependence, however, can easily be described. 
Suppose $\zeta_1$ and~$\zeta_2$ are two 
primitive $p$-th roots of unity and let~$\cH_1$ and~$\cH_2$ be the corresponding 
$I$-MacWilliams matrices. 
Then $\zeta_1^d=\zeta_2$ for some $0<d<p$ and, using the $\F_p$-linearity of~$\tau$, it 
is easy to check that $\cH_2=\cP(dI)\cH_1=\cH_1\cP(d^{-1}I)$. 
Making use of Remark~\ref{homogeneous}(a) this results in
$\cH_2\Lambda^t\cH_2^{-1}=\cH_1\Lambda^t\cH_1^{-1}$ and 
$\cH_2\Lambda\cH_2=\cH_1\Lambda\cH_1$.
Since all later expressions will be of either of these forms, our results later on do not depend on the choice of~$\zeta$. 
\end{remark}
%%%%%%%%%%%%%%%%%%%%%%%%%%%%%%%%%%

Obviously, the matrix~$\cH$ is symmetric.
Moreover, all MacWilliams matrices are invertible since the~$q^{\delta}$ different characters are linearly independent in the vector space of $\C$-valued functions on~$\F^\delta$. 
However, the inverse of these matrices can even easily be calculated.
Recall the matrices $\cP(P)$ from Definition~\ref{D-PP}.
%%%%%%%%%%%%%%%%%%%%%%%%%%%%%%
\begin{lemma} \label{Macw}
One has $\cH^2=\mathcal P(-I)$ and hence $\cH^4=I$.
Furthermore, 
\[
  \cH(P)=\cP(P)\cH=\cH\cP\big((P^t)^{-1}\big)\text{ for all }P\in \text{GL}_\delta(\F).
\]
In particular the inverse of a MacWilliams matrix is a MacWilliams matrix again. 
\end{lemma}
%%%%%%%%%%%%%%%%%%%%%%%%%%%%%%%%
\begin{proof}
For the computation of $\cH^2$ fix any pair $(X,Y)\in\cF$. Then, upon using the rules in
Remark~\ref{R-thetachar}(b) and~(c),
\[
  (\cH^2)_{X,Y}
              =q^{-\delta}\sum_{Z\in\F^{\delta}} \theta_X(Z)\theta_Z(Y)
      =q^{-\delta}\sum_{Z\in\F^{\delta}}\theta_{X+Y}(Z)
      =\!\left.\begin{cases}
           1,&\!\!\text{if }Y=-X\\ 0,&\!\!\text{else}
       \end{cases}
       \right\}\!=\cP(-I)_{X,Y}.
\]
The rest of the lemma can be checked in the same way using again Remark~\ref{R-thetachar}(c).
\end{proof}

%%%%%%%%%%%%%%%%%%%%%%%%%%%%%%%%%%%%%%%%
\begin{exa}\label{E-cH}
Let $p=q=2$ and $\delta=3$. Then $\zeta=-1$ and with respect to the lexicographic ordering\eqnref{e-lex} on $\F_2^3$ we obtain 
\[
   \cH=\frac{1}{\sqrt{8}}\left( 
   \begin{array}{rrrrrrrr}
     1&1&1&1&1&1&1&1\\ 1&-1&1&-1&1&-1&1&-1\\ 
     1&1&-1&-1&1&1&-1&-1\\1&-1&-1&1&1&-1&-1&1 \\ 
     1&1&1&1&-1&-1&-1&-1\\1&-1&1&-1&-1&1&-1&1 \\ 
     1&1&-1&-1&-1&-1&1&1\\ 1&-1&-1&1&-1&1&1&-1
   \end{array}\right).
\]
\end{exa}
%%%%%%%%%%%%%%%%%%%%%%%%%%%%%%%%%%%%%%%%%%%%%%%%

%%%%%%%%%%%%%%%%%%%%%%%%%%%%%%%%
\begin{remark}\label{R-MacWMat}
It should be mentioned that the MacWilliams matrices as presented here appear already in classical block code theory in the context of complete weight enumerators. 
 Given a block code $ C\subseteq\F^n$ the complete weight enumerator is defined as 
\[
  \cwe(C):=\sum_{(c_1,\ldots,c_n)\in C\;}\prod_{i=1}^nX_{c_i}\in \C[X_a\mid a\in \F].
\]
Obviously, we obtain the ordinary weight enumerator $\we(C)$ from $\cwe(C)$ by putting
$X_0=1$ and $X_a=W$ for all $a\in \F^*$.
Let now $\delta=1$ and $\cH\in \C^{q\times q}$ be the corresponding MacWilliams matrix.
Then~$\cH$ is the standard matrix interpretation of the $\C$-vector space automorphism
$h:\;\spann{X_a\mid a\in\F}_{\C}\longrightarrow \spann{X_a\mid a\in\F}_{\C}$ defined via 
$h(X_a)=q^{-\frac{1}{2}}\sum_{b\in\F}\zeta^{\tau(ab)}X_b$.
Extending~$h$ to a $\C$-algebra-homomorphism on $\C[X_a\mid a\in \F]$ it is well-known 
\cite[Ch.~5.6, Thm.~10]{MS77} that the complete weight enumerators of a $k$-dimensional 
block code $C\subseteq\F_q^n$ and its dual satisfy the MacWilliams identity $\cwe(C^{\perp})=q^{-k+\frac{n}{2}}h\big(\cwe(C)\big).$
At this point it is not clear to us why the MacWilliams matrix appears in the seemingly unrelated contexts of complete weight enumerators for block codes and adjacency matrices for convolutional codes. 
\end{remark}
%%%%%%%%%%%%%%%%%%%%%%%%%%%%%%%%%%%%%%

In the next section we will investigate a conjecture concerning a MacWilliams Identity Theorem for the adjacency matrices of convolutional codes and their duals. 
It states that for the data as in General Assumption~\ref{A-dataC} and for any 
$P\in \text{GL}_{\delta}(\F)$ the matrix $q^{-k}\HH\big(\cH(P)\Lambda(G)^t\cH(P)^{-1}\big)$ is a representative of the generalized adjacency matrix of~$\widehat{\cC}$ 
(in the sense of Remark~\ref{homogeneous}(b)), see Conjecture~\ref{C-MacWD}.  
Using Lemma~\ref{Macw} and the fact $\cH^t=\cH$ one easily observes that
$\cH\Lambda(G)\cH=\cP(-I)\big(\cH\Lambda(G)^t\cH^{-1}\big)^t$. 
Therefore the matrix $\cH\Lambda(G)\cH$ will be particularly helpful and will be 
studied first.
Let, as usual, the data be as in General Assumption~\ref{A-dataC} and Definition~\ref{Deflambda} and remember~$\widehat{r}$ from Proposition~\ref{duality}.
Put
\begin{equation}\label{e-ell}
   \ell_{X,Y}:=\big(\cH\Lambda(G)\cH\big)_{X,Y}\text{ for  }(X,Y)\in\cF.
\end{equation}
The entries $\ell_{X,Y}$ can be described explicitly. 
In the sequel we will use for any pair $(X,Y)\in\cF$ the short notation $(X,Y)^{\perp}:=\spann{(X,Y)}^{\perp}$ to denote the orthogonal space in~$\cF$.
The following result will be crucial for the MacWilliams Identity Conjecture as studied in the next section.
%%%%%%%%%%%%%%%%%%%%%%%%%%%%%%%%%%%%
\begin{theorem} \label{entriesl} 
Let $(X,Y)\in\cF$. Then 
\[
  \ell_{X,Y}=\begin{cases}0, &\text{if } (X,Y)\notin(\ker\Phi)^\bot,\\[1ex] 
          q^{-\widehat r}\we(\CC), &\text{if } (X,Y)\in\Delta^\bot,\\[1ex] 
	  \frac 1{q^{\delta}(q-1)}\Big(q\sum_{(Z_1,Z_2)\in (X,Y)^\bot}
	      \lambda_{Z_1,Z_2}-q^{\delta-\widehat r}\we(\CC)\Big) &\text{else.}
	     \end{cases}
\]
Furthermore, 
\[
    \ell_{X+U,Y+V}=\ell_{X,Y}\text{ for all }(U,V)\in\Delta^{\perp}.
\]
\end{theorem}
%%%%%%%%%%%%%%%%%%%%%%%%%%%%%%%%%%%%%%
The last statement can be regarded as a counterpart to Proposition~\ref{linearity}(c).
In fact, both these invariance properties will be needed to derive a correspondence between the matrix $\cH\Lambda(G)\cH$ and the adjacency matrix of the dual code in Section~\ref{S-Duality}.
\\[1ex]
\begin{proof}
Fix $(X,Y)\in\cF$. 
\\
1) We begin with proving the identity
\begin{equation}\label{lXY}
   q^{\delta}\ell_{X,Y}
   =\sum_{(Z_1,Z_2)\in (X,Y)^\bot}\lambda_{Z_1,Z_2}-\frac 1{q-1}
     \sum_{(Z_1,Z_2)\notin (X,Y)^\bot}\lambda_{Z_1,Z_2}.
\end{equation}
Using Remark~\ref{R-thetachar}(d), we have
\[
  q^{\delta}\ell_{X,Y}=\sum_{Z_1,\,Z_2\in\F^\delta}\theta_X(Z_1)\lambda_{Z_1,Z_2}\theta_{Z_2}(Y)
       =\sum_{(Z_1,Z_2)\in\cF}\theta_{(X,Y)}(Z_1,Z_2)\lambda_{Z_1,Z_2}.
\]
If $(X,Y)=(0,0)$, Equation\eqnref{lXY} follows. 
Thus let $(X,Y)\not=(0,0)$. 
Choose $(V_1,V_2)\in\cF$ such that $\mathcal F=(X,Y)^\bot\oplus\spann{(V_1,V_2)}$. This allows, recalling Remark~\ref{homogeneous}(a), to further simplify $\ell_{X,Y}$. Indeed, 
\begin{align*}
q^{\delta}\ell_{X,Y}
   &=\sum_{\alpha \in\F}\sum_{(Z_1,Z_2)\in (X,Y)^\bot}\theta_{(X,Y)}(Z_1+\alpha V_1,Z_2+\alpha V_2)
               \lambda_{Z_1+\alpha V_1,Z_2+\alpha V_2}\\
   &=\sum_{\alpha \in\F}\theta_{(X,Y)}(\alpha V_1,\alpha V_2)
     \sum_{(Z_1,Z_2)\in (X,Y)^\bot}\lambda_{Z_1+\alpha V_1,Z_2+\alpha V_2}\\
   &=\sum_{(Z_1,Z_2)\in (X,Y)^\bot}\lambda_{Z_1,Z_2}+
      \sum_{\alpha \in\F^*}\theta_{(X,Y)}(\alpha V_1,\alpha V_2)
      \sum_{(Z_1,Z_2)\in (X,Y)^\bot}\lambda_{Z_1+ V_1,Z_2+ V_2}.
\end{align*} 
Since $(X,Y)\not=(0,0)$ the character $\alpha\mapsto\theta_{(X,Y)}(\alpha V_1,\alpha V_2)$ is nontrivial on~$\F$ and thus
\begin{align}
  q^{\delta}\ell_{X,Y}
  &=\sum_{(Z_1,Z_2)\in (X,Y)^\bot}\lambda_{Z_1,Z_2}\;
    -\sum_{(Z_1,Z_2)\in (X,Y)^\bot}\lambda_{Z_1+V_1,Z_2+V_2} \label{e-ellXY}\\
  &=\sum_{(Z_1,Z_2)\in (X,Y)^\bot}\lambda_{Z_1,Z_2}\;
     -\frac 1{q-1}\sum_{(Z_1,Z_2)\notin (X,Y)^\bot}\lambda_{Z_1,Z_2}\;,\notag
\end{align}
where the last identity is again derived from Remark~\ref{homogeneous}(a) considering that $(X,Y)^\bot$ is a subspace of~$\cF$. 
This completes the proof of\eqnref{lXY}.
\\
2) Now we will prove each case of the second assertion separately. 
First let $(X,Y)\notin(\ker\Phi)^\bot$.
This implies $\ker\Phi\nsubseteq (X,Y)^\bot$. 
Hence $(V_1,V_2)$ above can be chosen in $\ker\Phi$ and therefore 
Proposition~\ref{linearity}(c) along with\eqnref{e-ellXY} yields $\ell_{X,Y}=0$. 
\\
3) Let $(X,Y)\in\Delta^\bot$, which implies $\Delta\subseteq (X,Y)^\bot$ and therefore $(\cF\backslash(X,Y)^\bot)\cap \Delta=\emptyset$. 
Using Equation\eqnref{lXY} together with the fact that $\lambda_{Z_1,Z_2}=0$ for $(Z_1,Z_2)\not\in\Delta$ one gets
\[
  q^\delta\ell_{X,Y}
   =\hspace*{-.6em}\sum_{(Z_1,Z_2)\in (X,Y)^\bot\cap\Delta}\hspace*{-.8em}\lambda_{Z_1,Z_2}
     -\frac 1{q-1}\sum_{(Z_1,Z_2)\in (\cF\backslash(X,Y)^\bot)\cap\Delta}
        \hspace*{-.6em}\lambda_{Z_1,Z_2}
   =\!\!\sum_{(Z_1,Z_2)\in\Delta}\lambda_{Z_1,Z_2}=q^{\delta-\widehat{r}}\we(\CC),
\]
where the last identity follows from Proposition~\ref{sumoverlambda}. 
\\
4) Finally, for the last case let $(X,Y)\in(\ker\Phi)^\bot\backslash\Delta^\bot$. 
Thus, $\ker\Phi\subseteq(X,Y)^\bot$, but $\Delta \nsubseteq (X,Y)^\bot$. 
Again, using Equation\eqnref{lXY} together with Proposition~\ref{sumoverlambda}, we obtain
\begin{align*}
  \ell_{X,Y}
   &=\frac 1{q^{\delta}(q-1)}\Big(
      (q-1)\sum_{(Z_1,Z_2)\in (X,Y)^\bot}\lambda_{Z_1,Z_2}
           -\sum_{(Z_1,Z_2)\notin (X,Y)^\bot}\lambda_{Z_1,Z_2}\Big)\\
   &=\frac 1{q^{\delta}(q-1)}\Big(
       q\sum_{(Z_1,Z_2)\in (X,Y)^\bot}\lambda_{Z_1,Z_2}-q^{\delta-\widehat r}\we(\CC)\Big).
\end{align*}
5) It remains to show $\ell_{X+U,Y+V}=\ell_{X,Y}$ for any $(U,V)\in\Delta^{\perp}$. 
Since $\Delta^{\perp}\subseteq(\ker\Phi)^{\perp}$
the statement is obvious in the first two cases of $\ell_{X,Y}$.
For the remaining case notice that in the expression for $\ell_{X,Y}$ we have
\[
   \sum_{(Z_1,Z_2)\in(X,Y)^{\perp}}\lambda_{Z_1,Z_2}
   =\sum_{(Z_1,Z_2)\in(X,Y)^{\perp}\cap\Delta}\lambda_{Z_1,Z_2}
   =\sum_{(Z_1,Z_2)\in(X+U,Y+V)^{\perp}\cap\Delta}\lambda_{Z_1,Z_2}
\]
for any $(U,V)\in\Delta^{\perp}$. This completes the proof. 
\end{proof}

At this point it is possible to derive a formula for the MacWilliams transformation~$\HH$ as defined in\eqnref{e-h} applied to the entries $\ell_{X,Y}$. 
It will play a central role in the next section. 
Recall from\eqnref{e-Cperpdef} the notation $\widehat\cC$ for the dual code of~$\cC$. 
Notice from\eqnref{e-ell} that the polynomials $\ell_{X,Y}$ are in $\C[W]_{\leq n}$ so that indeed the mapping~$\HH$ can be applied. 
%%%%%%%%%%%%%%%%%%%%%%%%%%%%%%%%%%%%%%%%%%%%%%%%%
\begin{proposition} \label{macw} 
Let $(X,Y)\in\cF$. Then 
\[
  q^{-k}\HH(\ell_{X,Y})
  =\begin{cases}
      0, &\text{if } (X,Y)\notin(\ker\Phi)^\bot,\\[1ex] 
      \we(\cCperpconst), &\text{if } (X,Y)\in\Delta^\bot, \\[1ex] 
      \frac 1{q-1}
        \Big(\we\big(\spann{\cCperpconst,\, c(X,Y)}\big)-\we\big(\cCperpconst\big)\Big),&\text{else,}
   \end{cases}
\] 
where in the last case $c(X,Y)$ is any element in  $\Big[\varphi\big((X,Y)^{\perp}\cap\Delta^*\big)+\cCconst\Big]^{\perp}\backslash\;\cCperpconst$.
\end{proposition}
%%%%%%%%%%%%%%%%%%%%%%%%%%%%%%%%%%%%%%%%%%%%%%%%%%
\begin{proof}
Use the form of $\ell_{X,Y}$ as given in Theorem~\ref{entriesl}. 
The first case is immediate as $\HH(0)=0$. 
The second case is exactly Corollary~\ref{C-MacWCC}.
The third case requires more work.
Thus, let $(X,Y)\in(\ker\Phi)^{\perp}\backslash\;\Delta^{\perp}$, hence 
$\ker\Phi\subseteq(X,Y)^\bot$, but $\Delta \nsubseteq (X,Y)^\bot$.
As a consequence, $(X,Y)^\bot \cap \Delta$ is a hyperplane of $\Delta$ and 
$\ker\Phi$ is contained in $(X,Y)^\bot \cap \Delta$.
Using the direct complement $\Delta^*$ of $\ker\Phi$ in~$\Delta$ as introduced in  Proposition~\ref{linearity} we obtain 
$(X,Y)^\bot \cap \Delta=\big((X,Y)^\bot \cap \Delta^*\big)\oplus\ker\Phi$ and 
$(X,Y)^\bot \cap \Delta^*$ is a hyperplane in $\Delta^*$. 
With the help of Proposition~\ref{linearity}(c) and Lemma~\ref{hom}(b) we get 
\[
  \sum_{(Z_1,Z_2)\in(X,Y)^{\perp}}\lambda_{Z_1,Z_2}
   =\sum_{(Z_1,Z_2)\in(X,Y)^{\perp}\cap\Delta}\lambda_{Z_1,Z_2}
    =q^{\delta-\widehat{r}}\sum_{(Z_1,Z_2)\in(X,Y)^{\perp}\cap\Delta^*}\lambda_{Z_1,Z_2}.
\]
By Lemma~\ref{weightenum} $\lambda_{Z_1,Z_2}=\we\big(\varphi(Z_1,Z_2)+\cCconst\big)$ and these cosets are pairwise disjoint for $(Z_1,Z_2)\in(X,Y)^{\perp}\cap\Delta^*$, see Corollary~\ref{C-Deltastar}. 
Therefore we obtain 
\[
  {\DS\sum_{(Z_1,Z_2)\in(X,Y)^{\perp}}\lambda_{Z_1,Z_2}
   =q^{\delta-\widehat{r}}\sum_{(Z_1,Z_2)\in(X,Y)^{\perp}\cap\Delta^*}
               \we\big(\varphi(Z_1,Z_2)+\cCconst\big)
    =q^{\delta-\widehat{r}}\we\big(H(X,Y)\big)} 
\]
where 
\[
       H(X,Y):=\!\!\bigcup_{(Z_1,Z_2)\in(X,Y)^{\perp}\cap\Delta^*}\!\!
                       \big(\varphi(Z_1,Z_2)+\cCconst\big)
                =\varphi\big((X,Y)^{\perp}\cap\Delta^*\big)+\cCconst.
\]
We will show next that $H(X,Y)^{\perp}=\spann{\cCperpconst,\,c(X,Y)}$ for some element $c(X,Y)$. 
In order to do so we need to compute the dimension of $H(X,Y)$. 
Since $\ker\varphi\cap\Delta^*=\{0\}$ we have 
$\dim\varphi\big((X,Y)^{\perp}\cap\Delta^*\big)
  =\dim\big((X,Y)^{\perp}\cap\Delta^*\big)=\dim\Delta^*-1$. 
Furthermore, Lemma~\ref{hom}(a) shows that $\varphi(\Delta^*)\cap\cCconst=\{0\}$. 
As a consequence, 
\[
  \dim H(X,Y)=\dim\Delta^*-1+\dim\cCconst=k+\widehat{r}-1.
\]
This implies $\dim H(X,Y)^{\perp}=n-k-\widehat r+1 =\dim\cCperpconst+1$. 
Furthermore, $H(X,Y)\subseteq\im\varphi+\cCconst=\CC$ and along with Proposition~\ref{duality}
this yields $\cCperpconst\subseteq H(X,Y)^{\perp}$. 
All this shows that there exists some $c(X,Y)\in H(X,Y)^{\perp}\backslash\;\cCperpconst$ such that 
\[
   H(X,Y)^{\perp}=\spann{\cCperpconst,\,c(X,Y)}.
\]
Now we can compute $q^{-k}\HH(\ell_{X,Y})$. From Theorem~\ref{entriesl} we derive
\begin{align*}
  q^{-k}\HH(\ell_{X,Y})
  &=q^{-k}\HH\Big(\frac 1{q^{\delta}(q-1)}
    \big(
      q\sum_{(Z_1,Z_2)\in (X,Y)^\bot}\lambda_{Z_1,Z_2}-q^{\delta-\widehat r}\we(\CC)\big)\Big)\\
  &=\frac 1{q-1}q^{-\widehat r-k}\Big(q \HH\big(\we\big(H(X,Y)\big)\big)
                                       -\HH\big(\we(\CC)\big)
				 \Big)\\
  &=\frac{q^{-\widehat r-k}}{q-1}\Big(q\cdot q^{k+\widehat r-1}
       \we\big(\spann{\cCperpconst,\,c(X,Y)}\big)-q^{k+\widehat r}\we(\cCperpconst)\Big),
\end{align*}
where the last identity is again due to\eqnref{e-MacWBC} and Corollary~\ref{C-MacWCC}.
This proves the desired result.
\end{proof}

The last proposition together with Lemma~\ref{weightenum} reveals an immediate resemblance of the entries $q^{-k}\HH(\ell_{X,Y})$ to that of any given adjacency matrix of the dual code of $\cC$. 
Indeed, firstly notice that both matrices have the same number of zero entries since 
$\#(\ker\Phi)^{\perp}=q^{\delta+\widehat r}$ is exactly the number of connected state pairs of the dual code. 
Moreover, Proposition~\ref{macw} tells us that $q^{-k}\HH(\ell_{X,Y})$ has $\#\Delta^{\perp}=q^{\delta-r}$ entries equal to $\we(\cCperpconst)$.
Applying Lemmas~\ref{weightenum} and~\ref{hom} to the dual code~$\widehat{\cC}$ we see 
that the adjacency matrix of the dual code has the same number of entries equal to
$\we(\cCperpconst)$.
The remaining entries also have an analogous form. 
All this indicates that there might be a strong relation between $q^{-k}\big(\HH(\ell_{X,Y})\big)$ and the adjacency matrix of the dual code. 
This will be formulated in a precise conjecture in the next section and proven for a specific class of codes. 
The difficulty for proving this will be, among other things, that we need a concrete description of the mapping
$(\ker\Phi)^\bot\backslash\;\Delta^\bot\longrightarrow \cF$, $(X,Y)\mapsto c(X,Y)$ as used in the last part of Proposition~\ref{macw}.

%%%%%%%%%%%%%%%%%%%%%%%%%%%%%%%%%%%%%%%%%%%%%
\section{A MacWilliams Identity for Convolutional Codes}\label{S-Duality}
\setcounter{equation}{0}
%%%%%%%%%%%%%%%%%%%%%%%%%%%%%%%%%%%%%%%%%%%%%%
In this section we will formulate the MacWilliams identity and prove it for a particular class of codes. 
Let again the data be as in\eqnref{e-Fq} and General Assumption~\ref{A-dataC}. Denote the associated adjacency matrix $\Lambda(G)$ simply by~$\Lambda$. 
Furthermore, let $\widehat\cC$ be the dual code. 
We fix the following notation. 
%%%%%%%%%%%%%%%%%%%%%%%%%%%%%%%%%%%%%%%%%%%%%%%%%
\begin{ass}\label{A-dataperp}
Let $\widehat\cC$ have encoder matrix $\widehat G\in\F[z]^{(n-k)\times n}$ and let 
the corresponding controller canonical form be denoted by $(\widehat A,\widehat B,\widehat C, \widehat D)$. 
Moreover let the associated adjacency matrix be written as
$\widehat\Lambda=:\big(\widehat\lambda_{X,Y}\big)$ and let $\widehat\Delta$ be the space of connected state pairs for $\widehat\cC$. Finally, we define the mappings 
$\widehat\varphi$ and $\widehat\Phi$ for the code~$\widehat\cC$ analogously to\eqnref{e-phi} and Lemma~\ref{hom} and, the spaces $\widehat\Delta^-$ and 
$\widehat\Delta^*$ analogously to Proposition~\ref{linearity}.
Recall from by Proposition~\ref{duality} that $\widehat\cC$ has $\widehat r$ nonzero Forney indices.
\end{ass}
%%%%%%%%%%%%%%%%%%%%%%%%%%%%%%%%%%%%%%
We know from\eqnref{e-Cperp} that~$\cC$ and~$\widehat\cC$ both have degree~$\delta$ and thus the adjacency matrices $\Lambda$ and $\widehat\Lambda$ are both in $\C[W]^{q^{\delta}\times q^{\delta}}$.
As a consequence we have all results of Section~\ref{S-adjmatrix} literally available in a $\widehat{\ \,}$ -version, and we will make frequent use of them. 

Notice that duality implies $G\widehat{G}^t=0$. From Remark~\ref{R-CCF} we know that
\[
   G(z)=B(z^{-1}I-A)^{-1}C+D,\quad \widehat{G}(z)=\widehat{B}(z^{-1}I-\widehat{A})^{-1}\widehat{C}+\widehat{D}.
\]
Since $D,\,\widehat D$ both have full row rank this implies 
\begin{equation}\label{e-DDhat}
    \im D=\ker\widehat{D}^t.
\end{equation}

Now we can formulate our conjecture. 
Recall the definition of the MacWilliams matrix~$\cH$ from Definition~\ref{D-MacWMat}.
%%%%%%%%%%%%%%%%%%%%%%%%%%%%%%%%%%%%%%%%%%%%%%
\begin{conj}\label{C-MacWD}
The matrix  $q^{-k}\HH\big(\cH\Lambda^t\cH^{-1}\big)$, where~$\HH$ is applied entrywise 
to the given matrix, is a representative of the generalized adjacency matrix 
of~$\widehat\cC$.
In other words, there exists some $P\in \text{GL}_{\delta}(\F)$ such that 
\begin{equation}\label{e-MacWConj}
   \widehat\lambda_{X,Y}=q^{-k}\HH\big((\cH\Lambda^t\cH^{-1})_{XP,YP}\big)
   \text{ for all }(X,Y)\in\cF.
\end{equation}
\end{conj}
%%%%%%%%%%%%%%%%%%%%%%%%%%%%%%%%%%%%%%%%%%%%%%%%%%%
Recall from Remark~\ref{homogeneous}(b) that the adjacency matrices for two different minimal encoders of~$\widehat{\cC}$ differ by conjugation with a suitable matrix $\cP(P)\in\Pi$.
This explains the presence of the matrix $P\in \text{GL}_{\delta}(\F)$ above.
Of course,~$P$ depends on the chosen encoders~$G$ and~$\widehat{G}$.
It is worth mentioning that in the case $\delta=0$ Identity\eqnref{e-MacWConj} immediately 
leads to the MacWilliams identity for block codes as given in\eqnref{e-MacWBC}.
Notice also that, due to Lemma~\ref{Macw} and Equation\eqnref{e-Pitrafo}, the conjecture implies the same statement if we replace~$\cH$ by an arbitrary $Q$-MacWilliams 
matrix~$\cH(Q)$.

The conjecture is backed up by many numerical examples.
A proof, however, is still open for the general case.
As a first step a somewhat weaker result will be proven in Theorem~\ref{isom}.
Thereafter we will fully prove the conjecture for codes where $\delta=\widehat{r}$ or $\delta=r$. 
In that case we will even be able to precisely tell which transformation matrix 
$P\in \text{GL}_{\delta}(\F)$, depending on~$G$ and~$\widehat{G}$, to choose 
for\eqnref{e-MacWConj} to be true. 
We need the following lemma. 
It still applies to the general situation.

%%%%%%%%%%%%%%%%%%%%%%%%%%%%%%%%%%%%%%%%%%%%%%%%
\begin{lemma} \label{constructf}
Let 
\[
   M:=\begin{pmatrix}\widehat C C^t& \widehat C (B^t D)^t\\\
      \widehat B^t\widehat D C^t&0
      \end{pmatrix}\in \F^{2\delta\times 2\delta}.
\] 
Then 
\begin{alphalist}
\item $\im M\subseteq(\ker\Phi)^\bot$.
\item $\ker \widehat{\Phi}\oplus \widehat \Delta^-\subseteq\ker M$.
\item $\im M\cap\Delta^{\perp}=\{0\}$.
\item $M$ is injective on $\widehat \Delta ^*$.
\item $\rank M=r+\widehat r$ and $\cM\oplus\Delta^{\perp}=(\ker\Phi)^\bot$
      where$\;\cM:=\!\im M\!=\{(X,Y)M\,|\,(X,Y)\in\widehat\Delta^*\}$. 
\end{alphalist}
\end{lemma}
%%%%%%%%%%%%%%%%%%%%%%%%%%%%%%%%%%%%%%%%%%%%%%%

\begin{proof}
First notice that by\eqnref{e-DDhat} 
$M^t=\Smalltwomat{C}{B^tD}\big(\widehat{C}^t\ (\widehat{B}^t\widehat{D})^t)$ 
and therefore    
\begin{equation}\label{e-betaM}
   \beta\big((X',Y'),\,(X,Y)M\big)=(X' ,Y' )M^t (X,Y)^t
   =\beta\big(\varphi(X' ,Y' ),\,\widehat\varphi(X,Y)\big)
\end{equation}
for all $(X,Y),\,(X',\,Y')\in\cF$.
Remember also that $\widehat\varphi(X,Y)\in\CCperp$ for all $(X,Y)\in\cF$.
\\[.5ex]
(a) follows from\eqnref{e-betaM} since for $(X',Y')\in\ker\Phi$ we have $\varphi(X',Y')\in\cCconst=(\CCperp)^{\perp}$.
\\[.5ex]
(b) If $(X,Y)\in\ker\widehat\Phi\oplus\widehat{\Delta}^-$, then 
$\widehat\varphi(X,Y)\in\cCperpconst$ by Corollary~\ref{C-Deltastar} and Lemma~\ref{hom}(a).
Thus $\widehat\varphi(X,Y)\in(\CC)^{\perp}$ while $\varphi(X',Y')\in\CC$ for all $(X',Y')\in\cF$.
Now\eqnref{e-betaM} along with the regularity of the bilinear form~$\beta$ shows 
$(X,Y)M=(0,0)$. 
\\[.5ex]
(c) Let $(X,Y)M\in\Delta^{\perp}$. 
Then by\eqnref{e-betaM} we have $\widehat\varphi(X,Y)\in\varphi(\Delta)^{\perp}$.
Since also $\widehat\varphi(X,Y)\in\CCperp=(\cCconst)^{\perp}$, we obtain from\eqnref{e-imphi}
and Proposition~\ref{duality} that $\widehat\varphi(X,Y)\in\cCperpconst$.
But then $(X,Y)\in\ker\widehat\Phi$ and~(b) implies $(X,Y)M=(0,0)$.
\\[.5ex]
(d) 
Let $(X,Y)M=0$ for some $(X,Y)\in\widehat\Delta^*$. 
Similarly to~(c) we obtain by use of\eqnref{e-betaM} and\eqnref{e-imphi}
\[
  \widehat\varphi(X,Y)\in(\im\varphi)^{\perp}\cap\CCperp
  =(\im\varphi)^{\perp}\cap(\cCconst)^{\perp}=(\im\varphi+\cCconst)^{\perp}
  =\cCperpconst.
\]
But this means that $(X,Y)\in\ker\widehat{\Phi}$ and the assumption 
$(X,Y)\in\widehat\Delta^*$ finally yields $(X,Y)=(0,0)$.
\\[.5ex]
(e) The rank assertion follows from~(d) and~(b) since $\dim\widehat\Delta^*=r+\widehat{r}$ 
and $\dim(\ker\widehat\Phi\oplus\widehat\Delta^-)=2\delta-(r+\widehat r)$. 
The rest is immediate from the above and $\dim(\ker\Phi)^\bot-\dim\Delta^{\perp}=r+\hat r$.
\end{proof}

The following result will be crucial for investigating Conjecture~\ref{C-MacWD}.
%%%%%%%%%%%%%%%%%%%%%%%%%%%%%%%%%%%%%%
\begin{theorem}\label{T-elltrafo}
Let $M\in\F^{2\delta\times2\delta}$ be as in Lemma~\ref{constructf}.
Then 
\[
  \widehat\lambda_{X,Y}=q^{-k}\HH(\ell_{(X,Y)M})\text{ for all }(X,Y)\in \widehat\Delta.
\]
\end{theorem}
%%%%%%%%%%%%%%%%%%%%%%%%%%%%%%%%%%%%%%%
\begin{proof}
Recall that $\widehat{\Delta}=\ker\widehat\Phi\oplus\widehat{\Delta}^*$. 
For $(X',Y')\in\ker\widehat\Phi$ and $(X,Y)\in\widehat{\Delta}^*$ we have 
$\widehat\lambda_{X'+X,Y'+Y}=\widehat\lambda_{X,Y}$ due to Proposition~\ref{linearity}(c).
Furthermore,  
$\ell_{(X',Y')M+(X,Y)M}=\ell_{(X,Y)M}$ by Lemma~\ref{constructf}(b).
Hence it suffices to show the result for $(X,Y)\in\widehat{\Delta}^*$.
For $(X,Y)=(0,0)$ the result is obviously true by Lemma~\ref{weightenum} and Proposition~\ref{macw}.
Thus let $(X,Y)\not=(0,0)$.
By Lemma~\ref{constructf}(e) this yields $(X,Y)M\in(\ker\Phi)^{\perp}\backslash\;\Delta^{\perp}$.
Hence $q^{-k}\HH(\ell_{(X,Y)M})$ needs to be computed according to the last case in Proposition~\ref{macw}. 
In order to do so we need to find a vector $c\big((X,Y)M\big)$ satisfying the requirements given there. 
We will show that $\widehat\varphi(X,Y)$ is such a vector.
First of all, it is clear that $\widehat\varphi(X,Y)\in\CCperp=(\cCconst)^{\perp}$. 
Moreover, due to $(X,Y)\in\widehat\Delta^*\backslash\{0\}$ we have 
$\widehat\varphi(X,Y)\not\in\cCperpconst$.
Finally applying\eqnref{e-betaM} to $(X',Y')\in\big((X,Y)M\big)^{\perp}\cap\Delta^*$ 
shows that
$\widehat\varphi(X,Y)\in\Big[\varphi\Big(\big((X,Y)M\big)^{\perp}\cap\Delta^*\Big)\Big]^{\perp}$.
All this shows that we may choose $c\big((X,Y)M\big)$ in Proposition~\ref{macw}
as $\widehat\varphi(X,Y)$.
Now that proposition yields 
\[
  q^{-k}\HH(\ell_{(X,Y)M})=\frac{1}{q-1}
  \Big(\we\big(\spann{\widehat\varphi(X,Y),\,\cCperpconst}\big)-\we(\cCperpconst)\Big),
\]
and this coincides with $\widehat\lambda_{X,Y}$ due to Lemma~\ref{weightenum}.
\end{proof}

For the sequel let $\cG$ be any direct complement of $(\ker\Phi)^{\perp}$ in~$\cF$. 
Due to Lemma~\ref{constructf}(e) we have the following decompositions of~$\cF$.
\begin{equation}\label{e-diagram}
\begin{array}{l}
    \hspace*{4.4cm}\widehat\Delta\\[.2ex]
    \hspace*{2.8cm}\overbrace{\hspace*{3.5cm}}\\[-.3ex]
    \xymatrix{
	\cF\ar[d]_{f}&=\!\!&\widehat\Delta^*\ar[d]_{f_0}\!\!&\oplus\!\!\!\!&        \ker\widehat\Phi\ar[d]_{f_1}&\oplus&\widehat\Delta^-\ar[d]_{f_2}\\
%	&  &\ar[d]_{f_0}   &      &\ar[d]_{f_1}      &      & \ar[d]_{f_2} \\
	\cF\,&=&\!\!\cM             &\!\!\oplus\!\!\!\!\!\!&\Delta^{\perp}   &\!\!\oplus\!\!&\cG
     }\\[-1.8ex]
    \hspace*{2.8cm}\underbrace{\hspace*{3.5cm}}\\[1ex]
    \hspace*{3.9cm}(\ker\Phi)^{\perp}
\end{array}  
\end{equation}
where, due to identical dimensions, there exist isomorphisms in each column. 
For~$f_0$ we choose the isomorphism induced by the matrix~$M$ from Lemma~\ref{constructf}, 
and thus $\cM=\im M$ as before.
This picture leads to the following result. 
%%%%%%%%%%%%%%%%%%%%%%%%%%%%%%%%%%%%%%%
\begin{theorem} \label{isom}
Consider the diagram\eqnref{e-diagram} and let the isomorphism $f_0$ be induced by the matrix~$M$ from Lemma~\ref{constructf}. 
Fix any isomorphisms~$f_1$ and~$f_2$ in the diagram.
Let $f:=f_0\oplus f_1\oplus f_2$ be the associated automorphism on~$\cF$. Then
\begin{equation}\label{e-trafo1}
  \widehat\lambda_{X,Y}=q^{-k} \HH\big((\cH\Lambda\cH)_{f(X,Y)}\big)
  \text{ for all }(X,Y)\in\cF. 
\end{equation}
As a consequence,
\[
   \widehat\lambda_{f^{-1}(-Y,X)} =q^{-k}\HH\big((\cH\Lambda^t\cH^{-1})_{X,Y}\big)
   \text{ for all }(X,Y)\in\cF. 
\] 
In particular, the entries of the matrices $\widehat{\Lambda}$ and $q^{-k}\HH\big(\cH\Lambda^t\cH^{-1}\big)$ coincide up to reordering.
\end{theorem}
%%%%%%%%%%%%%%%%%%%%%%%%%%%%%%%%%%%% 
\begin{proof}
Recall from\eqnref{e-ell} that $(\cH\Lambda\cH)_{f(X,Y)}=\ell_{f(X,Y)}$. 
We have to consider three cases.
\\
1) If $(X,Y)\not\in\widehat\Delta$, then $f(X,Y)\not\in(\ker\Phi)^{\perp}$ and 
$\widehat\lambda_{X,Y}=0=q^{-k}\HH(\ell_{f(X,Y)})$ due to the very definition  of~$\widehat\Delta$ and Proposition~\ref{macw}.
\\
2) If $(X,Y)\in\ker\widehat\Phi$ then $\widehat\varphi(X,Y)\in\cCperpconst$ and 
$f(X,Y)\in\Delta^{\perp}$. Now Lemma~\ref{weightenum} as well as Proposition~\ref{macw}
yield $\widehat\lambda_{X,Y}=\we(\cCperpconst)=q^{-k}\HH(\ell_{f(X,Y)})$.
\\
3) For the remaining case we have $(X,Y)\in\widehat\Delta\backslash\;\ker\widehat\Phi$.
Writing $(X,Y)=(X_1,Y_1)+(X_2,Y_2)$ where $(X_1,Y_1)\in\widehat\Delta^*$ and $(X_2,Y_2)\in\ker\widehat\Phi$, Proposition~\ref{linearity}(c) yields $\widehat\lambda_{X,Y}=\widehat\lambda_{X_1,Y_1}$ while
Theorem~\ref{entriesl} implies $\ell_{f(X,Y)}=\ell_{(X_1,Y_1)M}$.
Now the result follows from Theorem~\ref{T-elltrafo}.
\\
For the second statement put $\Gamma:=\cH\Lambda^t\cH^{-1}$.
Notice first that Lemma~\ref{Macw} and the definition of $\cP(-I)$ as given in~\ref{D-PP} 
yield $\cH\Lambda\cH=\cP(-I)\Gamma^t$.
This implies $(\cH\Lambda\cH)_{X,Y}=\Gamma_{Y,-X}$. 
Now we obtain from\eqnref{e-trafo1} $\widehat\lambda_{f^{-1}(X,Y)}=q^{-k}\HH(\Gamma_{Y,-X})$
and thus $\widehat\lambda_{f^{-1}(-Y,X)}=q^{-k}\HH(\Gamma_{X,Y})$.
This concludes the proof.
\end{proof}

It needs to be stressed that the theorem does not prove Conjecture~\ref{C-MacWD} since 
we did not show that $f^{-1}(-Y,X)=(XQ,YQ)$ for some suitable 
$Q\in \text{GL}_{\delta}(\F)$ and all $(X,Y)\in\cF$.
The difficulty in proving Conjecture~\ref{C-MacWD} consists precisely in finding
isomorphisms $f_0,\,f_1,\,f_2$ for Diagram\eqnref{e-diagram} such that $f^{-1}(-Y,X)$
has such a form. 
This will be accomplished next for the class of convolutional codes for which either~$r$ 
or~$\widehat r$ is equal to~$\delta$.

We begin with the case where $\widehat r=\delta$. Notice that this is equivalent to saying that all nonzero Forney indices of~$\widehat\cC$ have value one. 
Therefore, in this case
\begin{equation}\label{e-Ghat}
   \widehat G=\begin{pmatrix}\widehat D_1\\\widehat D_2\end{pmatrix}
              +z\begin{pmatrix}\widehat C\\ 0\end{pmatrix}
   \text{ where }
   \widehat D=\begin{pmatrix}\widehat D_1\\\widehat D_2\end{pmatrix}
             =\begin{pmatrix}\widehat B^t\widehat D\\\widehat D_2\end{pmatrix}
   \text{ and }
   \rank\begin{pmatrix}\widehat C\\\widehat D_2\end{pmatrix}=n-k, 
\end{equation}
where the last part follows from minimality of the encoder~$\widehat G$.
Furthermore, $\im\widehat D_2=\cCperpconst$. 
We will need the following technical lemma.
%%%%%%%%%%%%%%%%%%%%%%%%%%%%%%%%%
\begin{lemma}\label{L-relccfd}
Let $\widehat r=\delta$. Then the controller canonical forms satisfy
\begin{arabiclist}
\item $\widehat{B}^t\widehat{D}C^t\in \text{GL}_{\delta}(\F)$, 
\item $\widehat C D^t B+\widehat C C^t A=-\widehat B^t\widehat D C^t$.
\end{arabiclist}
\end{lemma}
%%%%%%%%%%%%%%%%%%%%%%%%%%%%%%%%%%%%%%
\begin{proof}
(1) Since $\widehat{r}=\delta$ we have $\widehat{B}^t=(I_{\delta},0)$ and thus $\widehat{B}^t\widehat{D}$ consists of the first~$\delta$ rows of~$\widehat{D}$.
As a consequence, $\widehat{B}^t\widehat{D}$ has full row rank~$\delta$.
Suppose now that $\rank\widehat{B}^t\widehat{D}C^t<\delta$. 
Since $\widehat{D}D^t=0$ this implies $\rank\widehat{B}^t\widehat{D}(C^t,D^t)<\delta$.
Hence there exists a nonzero vector $a\in\F^{\delta}$ such that 
$a\widehat{B}^t\widehat{D}(C^t,D^t)=0$. 
In other words, $a\widehat{B}^t\widehat{D}\in(\CC)^{\perp}=\cCperpconst$, where the last identity follow from Proposition~\ref{duality}.
Now the full row rank of $\widehat{B}^t\widehat{D}$ shows that 
$\im\widehat{B}^t\widehat{D}\cap\cCperpconst\not=\{0\}$, a contradiction to Remark~\ref{R-CC}(2) applied to the code~$\widehat\cC$. 
Hence $\rank\widehat B^t\widehat D C^t=\delta$.
\\
(2) By duality we have $\widehat G G^t=0$. From Remark~\ref{R-CCF} we know that the 
controller canonical forms determine the corresponding encoders via 
\[
   G(z)=B(z^{-1}I-A)^{-1}C+D=B\sum_{l\geq1}z^lA^{l-1}C+D \text{ and }
   \widehat G(z)=\widehat B\sum_{l\geq1}z^l\widehat A^{l-1}\widehat C+\widehat D
\]
where due to nilpotency the sums are finite. 
Since $\widehat r=\delta$ all Forney indices of $\widehat\cC$ are at most one and therefore $\widehat A=0$. 
Hence $\widehat G=z\widehat B\widehat C+\widehat D$. Now we compute 
\[
  0=\widehat G G^t=z(\widehat D C^tB^t+\widehat B\widehat CD^t)
      +\sum_{l\geq 2}z^l\big(\widehat DC^t(A^t)^{l-1}B^t+\widehat B\widehat C C^t(A^t)^{l-2}B^t\Big).
\]
Hence the coefficients of $z^l,\,l\geq1$, are zero. 
Left multiplying the coefficient of~$z$ by $\widehat B^t$, right multiplying it by~$B$, and using $\widehat B^t\widehat B=I$ this implies the identity
\begin{equation}\label{e-equ1}
  \widehat B^t\widehat D C^tB^tB+\widehat CD^tB=0.
\end{equation}
Furthermore, if we multiply the coefficient of~$z^l$ by $\widehat B^t$ from the left and by $BA^{l-1}$ from the right we obtain
\begin{align*}
  0&=\big(\widehat B^t\widehat DC^t(A^t)^{l-1}B^t
      +\widehat C C^t(A^t)^{l-2}B^t\big)BA^{l-1}
    =\big(\widehat B^t\widehat DC^tA^t
      +\widehat C C^t\big)\big((A^t)^{l-2}B^tBA^{l-1}\big)\\
   &=\big(\widehat B^t\widehat DC^tA^t
      +\widehat C C^t\big)\big((A^t)^{l-2}(I-A^tA)A^{l-1}\big)
\end{align*}
where the last identity is due to Remark~\ref{remarkconcan}. Addition of these equations yields
\[
  0=\big(\widehat B^t\widehat DC^tA^t+\widehat C C^t\big)
      \sum_{l\geq 2}\big((A^t)^{l-2}A^{l-1}-(A^t)^{l-1}A^{l}\big)
   =\widehat B^t\widehat DC^tA^tA+\widehat C C^tA,
\]
where the last identity follows from the nilpotency of~$A$.
Now we conclude with Remark~\ref{remarkconcan} and the aid of\eqnref{e-equ1}
\[
  \widehat C C^tA=-\widehat B^t\widehat DC^tA^tA
  =-\widehat B^t\widehat DC^t+\widehat B^t\widehat DC^tB^tB
  =-\widehat B^t\widehat DC^t-\widehat CD^tB,
\]
which is what we wanted. 
\end{proof}
Now we can present an isomorphism~$f_1$ for Diagram\eqnref{e-diagram}. 

%%%%%%%%%%%%%%%%%%%%%%%%%%%%%%%%%%%%%%%%%%%%%%%%
\begin{lemma} \label{rgleichdelta}
Let $\widehat r=\delta$. Define 
\[
    M_1:=\begin{pmatrix}-\widehat CC^t&\widehat CC^tA\\0&0\end{pmatrix}
    \in\F^{2\delta\times2\delta}.
\]
Then
\begin{alphalist}
\item $\im M_1\subseteq \Delta^\bot$.
\item $\ker \widehat{\Phi}\cap\ker M_1=\{0\}$.
\item $\rank M_1=\delta-r$.
\end{alphalist}
As a consequence, $M_1$ induces an isomorphism 
$f_1:\ker\widehat{\Phi}\longrightarrow \Delta^\bot$.
\end{lemma}
%%%%%%%%%%%%%%%%%%%%%%%%%%%%%%%%%%%%%%%%%%%%%%
\begin{proof}
(a) 
Using Lemma \ref{Deltabot} it suffices to show that $(\widehat CC^t)_{ij}=0$ for all 
$1\leq i\leq \delta$ and $j\in\cJ$. 
From\eqnref{e-Ghat} we see that all rows in $\widehat C$ are leading coefficient rows in $\widehat G$. 
By definition of the controller canonical form the $j$th rows of~$C$, where $j\in\cJ$, are leading coefficient rows of the encoder~$G$. 
Now $\widehat G G^t=0$ implies the desired result. 
As a consequence we also have $\rank M_1\leq \delta-r$.
\\[.5ex]
(b) From Lemma~\ref{L-relccfd} we know that $\widehat B^t\widehat D C^t\in \text{GL}_{\delta}(\F)$.
Let now $(X,Y)\in\ker \widehat{\Phi}\cap\ker M_1$. 
Then $X\widehat CC^t=0$ and, due to Lemma~\ref{constructf}(b) we also have $(X,Y)M=0$. 
As a consequence, $X\widehat CC^t+Y\widehat B^t\widehat D C^t=Y\widehat B^t\widehat D C^t=0$ and from the above we conclude $Y=0$. 
Now $(X,0)\in\ker\widehat\Phi$ yields
$\widehat\varphi(X,0)=X\widehat C\in\cCperpconst$. 
Using the full row rank of the rightmost matrix in\eqnref{e-Ghat} we conclude $X=0$.
\\[.5ex]
(c) follows from~(a) and~(b) since $\dim\ker\widehat\Phi=\delta-r$.
\end{proof} 

Now we are able to prove our main result. 
The crucial step will be a suitable choice for the space $\widehat\Delta^*$.
Recall that, so far, $\widehat\Delta^*$ was just any direct complement of 
$\ker\widehat\Phi$ in $\widehat\Delta$. 
As we will see below, $\ker M_1$ is such a direct complement. 
%%%%%%%%%%%%%%%%%%%%%%%%%%%%%%%%%%%%%
\begin{theorem}\label{T-MacWDuality}
Let $\widehat r=\delta$, that is, each Forney index of the code~$\widehat{\cC}$ is at most~$1$. 
Then $Q:=-\widehat B^t\widehat D C^t\in \text{GL}_{\delta}(\F)$ and 
\[
  \widehat\lambda_{X,Y}=q^{-k}\HH\big((\cH\Lambda^t\cH^{-1})_{XQ,YQ}\big)
  \text{ for all }(X,Y)\in\cF.
\]
As a consequence, 
\begin{equation}\label{e-MM2}
   \widehat\Lambda=q^{-k}\HH\big(\cP(Q)\cH\Lambda^t\cH^{-1}\cP(Q)^{-1}\big),
\end{equation} 
where the MacWilliams transformation~$\HH$  has to be applied entrywise. 
\end{theorem}

%%%%%%%%%%%%%%%%%%%%%%%%%%%%%%%%%%%%%%
\begin{proof}
The invertibility of $Q=\widehat B^t\widehat D C^t$ has been shown in Lemma~\ref{L-relccfd}(1).
Choose the matrices~$M$ and~$M_1$ as in Lemmas~\ref{constructf} and~\ref{rgleichdelta}. 
By Lemma~\ref{L-relccfd}(2) we have
\begin{equation}\label{e-MM1}
   M+M_1=\begin{pmatrix} 0 &Q\\ -Q&0\end{pmatrix}\in \text{GL}_{2\delta}(\F).
\end{equation}
Notice that due to $\widehat r=\delta$ we have 
$\cF=\widehat\Delta$. 
In particular, the last column of Diagram~\ref{e-diagram} is trivial.
Lemma~\ref{rgleichdelta} shows that $\widehat\Delta^*:=\ker M_1$ is a direct complement of~$\ker\widehat\Phi$ in $\cF$.
Now define the automorphism $f:\cF\longrightarrow\cF$ as $f(X,Y)=(X,Y)(M+M_1)$. 
Using Lemma~\ref{constructf} and Lemma~\ref{rgleichdelta}, in particular
$\widehat\Delta^*=\ker M_1$ and $\ker\widehat\Phi\subseteq\ker M$,
we see that~$f$ is of the form $f=f_0\oplus f_1$ as required in Diagram\eqnref{e-diagram}.
Here~$f_0$ and~$f_1$ are induced by the matrices~$M$ and~$M_1$, respectively.
It is easy to see that $f^{-1}(-Y,X)=(XQ^{-1},YQ^{-1})$ and therefore
Theorem~\ref{isom} yields $\widehat\lambda_{XQ^{-1},YQ^{-1}}=q^{-k}\HH\big((\cH\Lambda^t\cH^{-1})_{X,Y}\big)$. 
Using\eqnref{e-Pitrafo} this implies the desired result. 
\end{proof}

%%%%%%%%%%%%%%%%%%%%%%%%%%%%%%%%%%%%%%%
\begin{exa} 
Note that the code~$\cC$ from Example~\ref{example1} satisfies $\widehat r=\delta$. 
Hence we can apply Theorem~\ref{T-MacWDuality} to this code. 
The automorphism $Q$ can be calculated as
$Q=\left(\begin{smallmatrix} 1&0&1\\1&0&0& \\0&1&0 \end{smallmatrix}\right)$ 
and, using the lexicographic ordering\eqnref{e-lex}, the permutation matrix is given by 
\[
  \cP(Q)=\begin{pmatrix}1&0&0&0&0&0&0&0\\0&0&1&0&0&0&0&0 \\
      0&0&0&0&1&0&0&0 \\0&0&0&0&0&0&1&0 \\
      0&0&0&0&0&1&0&0 \\0&0&0&0&0&0&0&1 \\
      0&1&0&0&0&0&0&0 \\0&0&0&1&0&0&0&0 
      \end{pmatrix}.
\]
Using this as well as the adjacency matrices given in Example~\ref{E-exa2} and the MacWilliams matrix~$\cH$ in Example~\ref{E-cH} one can check straightforwardly Identity\eqnref{e-MM2}.
\end{exa}
%%%%%%%%%%%%%%%%%%%%%%%%%%%%%%%%%%%%%%%%%%

We can easily transfer our result to convolutional codes with $\delta=r$.
%%%%%%%%%%%%%%%%%%%%%%%%%%%%%%%%%%%%%%
\begin{theorem} \label{T-MacWDuality2}
Let $\delta=r$, that is, each Forney index of the code~$\cC$ is at most~$1$.
Then $P:=-\widehat CD^tB \in \text{GL}_\delta(\F)$ and for all $(X,Y)\in\cF$ we have $\widehat\lambda_{X,Y}=q^{-k}\HH\big((\cH\Lambda^t\cH^{-1})_{XP,YP}\big)$. 
In other words,
\[
  \widehat\Lambda=q^{-k}\HH\big(\cP(P)\cH\Lambda^t\cH^{-1}\cP(P)^{-1}\big).
\]
\end{theorem}
%%%%%%%%%%%%%%%%%%%%%%%%%%%%%%%%%%%%%%
\begin{proof} 
Notice that we can apply Theorem~\ref{T-MacWDuality} to the code~$\widehat\cC$.
Hence $Q:=-B^tD\widehat C^t$ is regular and, since $\dim\widehat{\cC}=n-k$, we obtain 
$\Lambda=q^{-n+k}\HH(\cP(Q)\cH\widehat{\Lambda}^t\cH^{-1}\cP(Q)^{-1})$. 
Applying $\HH^2(f)=q^nf$ and the $\C$-linearity of~$\HH$ we arrive at
$q^{-k}\HH(\cH^{-1}\cP(Q)^{-1}\Lambda\cP(Q)\cH)=\widehat{\Lambda}^t$.
Transposing this equation and remembering that $\cH^t=\cH$ while $\cP(Q)^t=\cP(Q)^{-1}$, one gets
\[
  \widehat{\Lambda}=q^{-k}\HH(\cH\cP(Q)^{-1}\Lambda^t\cP(Q)\cH^{-1}).
\]
Now Lemma~\ref{Macw} yields $\cH\cP(Q)^{-1}=\cH\cP(Q^{-1})=\cP(Q^t)\cH=\cP(P)\cH$.
This concludes the proof.
\end{proof}

Incorporating the permutation matrix into the MacWilliams matrix, see Lemma~\ref{Macw}, we can state Theorems~\ref{T-MacWDuality} and~\ref{T-MacWDuality2} in terms of $P$-MacWilliams matrices only.
%%%%%%%%%%%%%%%%%%%%%%%%%%%%%%%%%%%%%%%%%%%
\begin{corollary}\label{C-closedform}\
\begin{arabiclist}
\item If $\widehat r=\delta$, then $Q=-\widehat B^t\widehat DC^t\in \text{GL}_\delta(\F)$ and
      $\widehat \Lambda=q^{-k}\HH\big(\cH(Q)\Lambda^t\cH(Q)^{-1}\big)$.
\item If $r=\delta$, then $P=-\widehat C D^tB\in \text{GL}_\delta(\F)$ and
      $\widehat \Lambda=q^{-k}\HH\big(\cH(P)\Lambda^t\cH(P)^{-1}\big)$.    
\end{arabiclist}
\end{corollary} 
%%%%%%%%%%%%%%%%%%%%%%%%%%%%%%%%%%%%%%%%

Using the notion of the generalized adjacency matrix as defined in Remark~\ref{homogeneous}(b) we obtain the following consequence, formulated independently of any chosen representation.

%%%%%%%%%%%%%%%%%%%%%%%%%%%%%%%%%%%%%%%%
\begin{theorem}
Let $r=\delta$ or $\widehat r=\delta$.
Then the generalized adjacency matrix of~$\cC$ uniquely determines the generalized adjacency 
matrix of the dual code~$\widehat{\cC}$. 
More precisely, let $[\Lambda]$ and~$[\widehat{\Lambda}]$ be the generalized adjacency matrices of~$\cC$ and~$\widehat{\cC}$, respectively. 
Then, in a suggestive notation,  
\[
   [\widehat{\Lambda}]=q^{-k}\HH(\cH[\Lambda]^t\cH^{-1}).
\]
\end{theorem}
%%%%%%%%%%%%%%%%%%%%%%%%%%%%%%%%%%%%%%%%

We close this section with an example supporting Conjecture~\ref{C-MacWD} that is not covered by the cases in Theorem~\ref{T-MacWDuality} or Theorem~\ref{T-MacWDuality2}.

%%%%%%%%%%%%%%%%%%%%%%%%%%%%%%%%%%%%%%%%
\begin{exa}\label{E-bigger}
Let $q=3$, thus $\F=\F_3$, and 
$G=\left(\begin{smallmatrix}1+z^2&2+z&0\\1&0&2\end{smallmatrix}\right)$.
Put $\cC=\im G$. 
Then~$G$ is a minimal basic matrix and thus~$\cC$ is a $(3,2,2)$ code.
The dual code is given by $\widehat{\cC}=\im\widehat{G}$ where 
$\widehat{G}=\begin{pmatrix}2+z& 2+2z^2&2+z\end{pmatrix}$. Notice that $r=\widehat{r}=1\not=\delta$. 
Using the controller canonical forms one can straightforwardly compute the adjacency matrices~$\Lambda,\,\widehat{\Lambda}\in\C[W]^{9\times9}$.
Then, via a systematic search one finds that Identity\eqnref{e-MacWConj} is satisfied if one chooses the regular matrix
$P=\left(\begin{smallmatrix}1&1\\1&2\end{smallmatrix}\right)\in \text{GL}_2(\F_3)$.
In the same way one can establish plenty of examples. 
\end{exa}
%%%%%%%%%%%%%%%%%%%%%%%%%%%%%%%%%%%%%%%%

%%%%%%%%%%%%%%%%%%%%%%%%%%%%%%%%%%%%%%%%%%%%%
\section{Unit Constraint-Length Codes}\label{S-AbGh}
\setcounter{equation}{0}
%%%%%%%%%%%%%%%%%%%%%%%%%%%%%%%%%%%%%%%%%%%%%
In the last section we want to have a closer look at codes with degree $\delta=1$, also called 
unit constraint-length codes or unit memory codes. 
Notice that in this case $r=\widehat{r}=\delta=1$. 
The situation now becomes particularly simple since, firstly, $\text{GL}_{\delta}(\F)=\F^*$ and, secondly, the adjacency matrices~$\Lambda$ and~$\widehat{\Lambda}$ do not depend on the 
choice of the encoder matrices~$G$ and~$\widehat{G}$. 
The latter is a consequence of Equation\eqnref{e-Pitrafo} along with Remark~\ref{homogeneous}(a),~(b).
Notice also that in Diagram\eqnref{e-diagram} the second and third column are trivial.
Using once more Remark~\ref{homogeneous}(a), we finally see that the statements of both Theorem~\ref{T-MacWDuality} and Theorem~\ref{T-MacWDuality2} reduce to the nice short 
formula
\begin{equation}\label{e-delta1}
    \widehat\Lambda=q^{-k}\HH(\cH\Lambda^t\cH^{-1}).
\end{equation}

In the paper~\cite{Ab92} the so-called weight enumerator state diagram has been studied for codes with degree one.  
They are defined as the state diagram of the encoder where each directed edge is labeled by the weight enumerator of a certain affine code. 
A type of MacWilliams identity has been derived for these objects~\cite[Thm.~4]{Ab92}.
It consists of a separate transformation formula for each of these labels.
After some notational adjustment one can show that the weight enumerator state diagram is in essence identical to the adjacency matrix of the code. 
Furthermore, if stated in our notation, the MacWilliams identity in~\cite[Thm.~4]{Ab92} reads as
\begin{equation}\label{e-AbGh}
  \widehat{\lambda}_{X,Y}=
    \begin{cases}
       q^{-k-1}\HH\big(\lambda_{0,0}+(q-1)(\lambda_{0,1}+\sum_{Y\in\F}\lambda_{1,Y})\big)
       &\text{if }(X,Y)=(0,0)\\[1ex]
      q^{-k-1}\HH\big(\lambda_{0,0}+q\lambda_{X,Y}-\lambda_{0,1}-\sum_{Y\in\F}\lambda_{1,Y}\big)
     &\text{else.}
    \end{cases}
\end{equation}
In the sequel we will briefly sketch that this result coincides with Identity\eqnref{e-delta1}.
In order to do so use again $\ell_{X,Y}$ as introduced in\eqnref{e-ell}. 
Then\eqnref{e-delta1} turns into
\begin{equation}\label{e-delta=1}
   \widehat{\lambda}_{X,Y}=q^{-k}\HH(\ell_{-Y,X})\text{ for all }(X,Y)\in\cF.
\end{equation}
Now we are in a position to derive\eqnref{e-AbGh}.
Consider first the case $(X,Y)=(0,0)$. 
Recalling Theorem~\ref{entriesl}, Proposition~\ref{sumoverlambda}, and 
Remark~\ref{homogeneous}(a) we find
\begin{align*}
   q\ell_{0,0}&=\we(\CC)
   =\sum_{(X,Y)\in\cF}\lambda_{X,Y}
   =\lambda_{0,0}+\sum_{Y\in\F^*}\lambda_{0,Y}+\sum_{X\in\F^*}\sum_{Y\in\F}\lambda_{X,Y}\\
   &=\lambda_{0,0}+(q-1)(\lambda_{0,1}+\sum_{Y\in\F}\lambda_{1,Y}).
\end{align*}
Using\eqnref{e-delta=1} this yields the first case of\eqnref{e-AbGh}.
For the second case let $(X,Y)\in\cF\,\backslash(0,0)$. 
Since $\Delta^{\perp}=\{0\}$ and $(\ker\Phi)^{\perp}=\cF$ one observes that 
in Theorem~\ref{entriesl} the third case has to be applied. 
Along with Proposition~\ref{sumoverlambda} and Remark~\ref{homogeneous}(a) this yields
\begin{align*}
  q\ell_{-Y,X}
    &=\frac 1{q-1}\Big(q\sum_{(Z_1,Z_2)\in (-Y,X)^\bot}\lambda_{Z_1,Z_2}-\we(\CC)\Big)
     =\frac 1{q-1}\Big(q\sum_{\alpha\in \F}\lambda_{\alpha X,\alpha Y}-\we(\CC)\Big)\\
   &=\frac 1{q-1}\Big(q(q-1)\lambda_{X,Y}+q\lambda_{0,0}-\sum_{(Z_1,Z_2)\in\cF}\lambda_{Z_1,Z_2}\Big)\\
  &=q\lambda_{X,Y}+\lambda_{0,0}
      -\frac 1{q-1}\sum_{(Z_1,Z_2)\in\cF\backslash \{(0,0)\}}\lambda_{Z_1,Z_2}
   =q\lambda_{X,Y}+\lambda_{0,0}-\lambda_{0,1}-\sum_{Y\in\F}\lambda_{1,Y}.
\end{align*}
Combining this with\eqnref{e-delta=1} leads to the second case of\eqnref{e-AbGh}. 

%%%%%%%%%%%%%%%%%%%%%%%%%%%%%%%%%%%%%%%%%%%%%%
\section*{Conclusion}
In this paper we studied the adjacency matrices for convolutional codes.
We introduced a transformation consisting of conjugation with the MacWilliams matrix followed by entrywise application of the MacWilliams Identity for block codes.
We proved that the resulting matrix coincides up to reordering of the entries with the adjacency matrix of the dual code, and we presented the reordering mapping explicitly.
This result can be regarded as a weak MacWilliams Identity for convolutional codes.
However, we strongly believe that the reordering of the entries can even be expressed in terms of an isomorphism on the state space, and indeed, we proved this statement for a particular class of convolutional codes. 
The general case has to remain open for future research.
%%%%%%%%%%%%%%%%%%%%%%%%%%%%%%%%%%%%%%%%%%%%%%%

%%%%%%%%%%%%%%%%%%%%%%%%%%%%%%%%%%%%%%%%%%%%%%%
\bibliographystyle{abbrv}
\bibliography{literatureAK,literatureLZ}
\end{document}